\documentclass[prd,aps,floatfix,amsmath,amssymb,twocolumn,nofootinbib]{revtex4-1}
\usepackage{amssymb,amsmath,amsthm,xcolor,graphicx}
\usepackage[margin=2cm]{geometry}
\usepackage[pdftex,breaklinks,colorlinks,
linkcolor=blue,
citecolor=teal,
anchorcolor=red,
urlcolor=cyan]{hyperref}

\begin{document}


\title{Simulations of gravitational collapse in null
  coordinates: III. Hyperbolicity}


\author{Carsten Gundlach}
\affiliation{Mathematical Sciences, University of Southampton,
  Southampton SO17 1BJ, United Kingdom} 

\date{25 April 2024}


\begin{abstract}

We investigate the well-posedness of the characteristic initial-boundary value problem for the Einstein equations in Bondi-like coordinates (including Bondi, double-null and affine). We propose a definition of strong hyperbolicity of a system of partial differential equations of any order, and show that the Einstein equations in Bondi-like coordinates in their second-order form used in numerical relativity do not meet it, in agreement with results of Giannakopoulos et al for specific first-order reductions. In the principal part, frozen coefficient approximation that one uses to examine hyperbolicity, we explicitly construct the general solution to identify the solutions that obstruct strong hyperbolicity. Independently, we present a first-order symmetric hyperbolic formulation of the Einstein equations in Bondi gauge, linearised about Schwarzschild, thus completing work by Frittelli. This establishes an energy norm ($L^2$ in the metric perturbations and selected first and second derivatives), in which the initial-boundary value problem, with initial data on an outgoing null cone and boundary data on a timelike cylinder or an ingoing null cone, is well-posed, thus verifying a conjecture by Giannakopoulos et al. Unfortunately, our method does not extend to the pure initial-value problem on a null cone with regular vertex. 

\end{abstract}


\maketitle
\tableofcontents


\section{Introduction}


Well-posedness of a system of partial differential equations (from now
on, PDEs) is defined as the existence and uniqueness of solutions and
their continuous dependence on the initial and boundary data.  For
formulations of the Einstein equations where surfaces of the time
coordinate $u$ are outgoing null cones, three PDE problems, among
others, are of interest: 1) the double null initial value problem,
with free data posed on an outgoing null cone $u=0$ and ingoing null
cone $v=0$ that intersect in a spacelike 2-sphere; 2) the
initial-boundary value problem, with free initial data posed on $u=0$
and boundary data on a timelike cylinder $r=r_0$, again intersecting
in a spacelike 2-sphere; and 3) the initial value problem on a null
cone $u=0$ with regular vertex $r=0$.

The mathematical literature has focused on the proof of well-posedness
of such problems in {\em some} coordinate system and formulation that
is convenient for the proof. This might be called geometric
well-posedness. By contrast, in numerical relativity, well-posedness
of the continuum problem is necessary for the existence of a stable
discretisation, but other considerations are equally important for the
choice of formulation of the Einstein equations, and so one wants a
proof of well-posedness in the formulation of choice.

We give two examples of this difference in emphasis. A famous proof of
geometric well-posedness of the Cauchy problem was given by
Four\`es-Bruhat in harmonic coordinates \cite{FouresBruhat1952}, but
these became useful for numerical relativity only through the
breakthrough work of Pretorius \cite{Pretorius2005}, which incorporated
two key modifications of lower-order terms: specific choices of the
``gauge source functions'' of \cite{Friedrich1985} that avoid coordinate
singularities, and the addition of ``constraint damping'' terms to
suppress violations of the harmonic and Einstein constraints arising
from numerical error \cite{constraintdamping}. In a second example,
much of the mathematical relativity literature uses double-null
coordinates on ingoing and outgoing null cones, but these are not
expected to be useful in numerical relativity beyond spherical
symmetry because the ingoing null cones form caustics. (It is one
purpose of the present series of papers to establish if outgoing null
cones also form caustics in specific strong-field applications.)

Geometric well-posedness for the double initial value problem was
established by Rendall \cite{Rendall1990} for smooth data, covering a
small neighbourhood of the intersection 2-sphere. The proof used
harmonic coordinates, made to coincide with $u$ and $v$ on the initial
surfaces $u=0$ and $v=0$.  This result was improved by Luk
\cite{Luk2011} for a larger region (small finite distances to the
future of the two initial null surfaces) and rougher data. The proof
used double-null coordinates, and estimates in $H^1$ of a null tetrad
and the Ricci rotation coefficients and curvature components in this
tetrad.  The initial value problem on a regular null cone is even harder
because of the need to characterise data near the tip that will give a
regular solution. Existence was proved by Chru\'sciel
\cite{Chrusciel2013}, see also \cite{ChruscielJezierski2011}.

In the present series of papers, we focus on class of coordinates
$(u,x,\theta,\varphi)$ introduced in
\cite{BondiBurgMetzner1962,Sachs1962}, and called ``Bondi-like'' in
\cite{GHZ2020}, where the surfaces of constant (retarded) time $u$ are
outgoing null cones and the lines of constant $(u,\theta,\varphi)$ are
their generators. (To avoid confusion about terminology, we mention
already that ``Bondi coordinates'' are further defined by the radial
coordinate $x$ being the area radius.)

In addition, we are interested in formulations of the Einstein
equations where a maximum number of equations can be solved as
ordinary differential equations (from now on, ODEs) along the
generators of the coordinate null cones, and only a minimum number,
namely two or three, contain $u$-derivatives: intuitively, the latter
are evolution equations for the two polarisations of gravitational waves,
plus, in double-null coordinates only, a third one for the area radius
$R$.

Such formulations are attractive for numerical relativity because they
are ``maximally constrained'', meaning that one solves the same
constraint equations on each time slice as on the initial time slice
(so that only free data are evolved), and because these constraints
are not elliptic equations but inhomogeneous first-order linear ODEs
that can be solved by integration. (An exception to this statement are
affine null coordinates, where one of the constraints is a homogeneous
linear second-order ODE.) 

Time evolutions beyond spherical symmetry on null cones emanating from
a regular centre have been carried out, for example, for supernova
core collapse \cite{Siebeletal2003} and for scalar field critical
collapse \cite{axinull_critscalar}, and vacuum evolutions using
Cauchy-characteristic matching along a timelike cylinder, for example in
\cite{Maetal2023}.

However, we are not aware of any previous well-posedness result
specifically in Bondi-like coordinates. In an incomplete attempt,
Frittelli \cite{Frittelli2005a} constructed a first-order symmetric
hyperbolic form of the vacuum Einstein equations in Bondi coordinates,
linearised about Minkowski space. An ``energy'' estimate, in $L^2$ of
the reduction variables, then follows. However, the perturbation
$\tilde V$ of the metric coefficient $V$ was omitted from the
system. This is possible in the linearisation about Schwarzschild
because $\tilde V$ couples to the other perturbations, but not vice versa.

The present paper is an attempt to reconcile the results of
\cite{Rendall1990,Luk2011,Chrusciel2013}, and the suggestive
incomplete result of \cite{Frittelli2005a}, with recent work of
Giannakopoulos and collaborators. Their paper \cite{GHZ2020} found
that a first-order reduction of the null cone formulation is weakly,
but not strongly, hyperbolic (and hence not symmetric hyperbolic),
that the lack of strong hyperbolicity is essentially a gauge problem
in any Bondi-like gauge \cite{GBHPZ2022}, and that it appears to
indeed break the convergence of numerical solutions with resolution,
both in toy models and in an open-source code for the null
initial-boundary value problem \cite{GBHPZ2023}.

The structure of the paper is as follows. As in the previous papers in
this series \cite{axinull_formulation, axinull_critscalar}, in
Sec.~\ref{section:hyperbolicity} we restrict to twist-free
axisymmetry, with a minimally coupled massless scalar field $\psi$ as
matter. We briefly restate the metric, the mathematical structure of
the field equations and the gauge choices we consider.  We then
linearise the equations, first around an arbitrary background and then
around Minkowski spacetime, drop lower-order terms, and ``freeze'' the
background coefficients by treating them as constants. We now have a
homogeneous system of linear PDEs with constant coefficients that we
shall call the ``toy model'' of the original system. The symbol of the
toy model is also the principal symbol of the original system, and
strong hyperbolicity is an algebraic property of this principal
symbol. 

In contrast to \cite{GHZ2020,Frittelli2005a}, we do not reduce
the PDEs to first order, but leave them in the form in which they are
solved numerically. We generalise the textbook definition of strong
hyperbolicity of first-order systems of PDEs to systems of PDEs of
arbitrary order. We show that, by this criterion, all known Bondi-like
null gauges are only weakly hyperbolic. We also find the general
solution of the toy model itself in closed form, and hence identify
the polynomial solutions that obstruct strong
hyperbolicity. (Appendix~\ref{appendix:toysystem} reminds the reader
of a textbook example of this phenomenon).

In Sec.~\ref{section:Frittelli}, we use a completely different
approach. We relax the restriction to twist-free axisymmetry, but
linearise about the Schwarzschild solution, and restrict to Bondi
gauge. We present a first-order symmetric hyperbolic form of the full
linearised equations, thus completing the result of
\cite{Frittelli2005a}. (We correct some minor errors in
Appendix~\ref{appendix:Frittellicorrections}.) To include all 6 metric
coefficients in the estimate, we need to include also 10 first and 7
second derivatives of the metric as variables in the system and hence
in the estimate. These are far from all first and second derivatives,
and their choice is crucial for symmetric hyperbolicity. To connect
with Sec.~\ref{section:hyperbolicity} with
Sec.~\ref{section:Frittelli}, we give the equivalent symmetric
hyperbolic form of the toy model in
Appendix~\ref{toymodelsymmetrichyperbolic}.

We then present well-posedness estimates for the initial value problem
on two intersecting null cones and the initial-boundary value problem
on a null cone intersecting a timelike world tube, closely following
\cite{Frittelli2004,Frittelli2005b}. Unfortunately, these methods do
not allow us to derive an estimate for the pure Cauchy problem on an
outgoing null cone with regular vertex, see
Appendix~\ref{appendix:tricks} for the technical obstructions.

We summarize and conclude in Sec.~\ref{section:conclusions}.


\section{Obstructions to strong hyperbolicity of the second-order form
of the Einstein equations in Bondi-like gauges}
\label{section:hyperbolicity}


\subsection{Metric and field equations in twist-free axisymmetry}


We begin with a brief review of our setup in this ection, see
\cite{axinull_formulation} for full details. We can write the metric
of any twist-free axisymmetric spacetime in the form
\begin{eqnarray}
\label{ymetric}
  ds^2&=&-2G\,du\,dx-H\,du^2 \nonumber \\
&&+R^2\left[e^{2S f}S^{-1}(dy+S\,b\,du)^2
+e^{-2S f}S\,d\varphi^2\right]. \nonumber \\ 
\end{eqnarray}
We use the angular coordinate $y:=-\cos\theta$, so that the range
$0\le\theta\le \pi$ corresponds to $-1\le y\le 1$, and the shorthand
$S:=1-y^2$. The other angular coordinate $\varphi$ has the usual range
$0\le\varphi<2\pi$. The Killing vector generating the axisymmetry is
$\partial_\varphi$. (We use the convention of equating vector fields
with derivative operators.) $(G,H,R,b,f)$ and the scalar field $\psi$
depend on $(u,x,y)$. In \cite{axinull_formulation} we assumed that
$R=0$ occurs at $x=0$ and is a regular centre, but we do not make this
assumption here.
 
Each surface ${\cal N}^+_u$ of constant $u$ is an outgoing null cone,
and is ruled by the null geodesics ${\cal L}^+_{u,y,\varphi}$, which
are also coordinate lines. Each surface ${\cal S}_{u,x}$ of constant
$u$ and $x$ is assumed to be spacelike, and has topology $S^2$ and
area $4\pi R^2$. The
outgoing future-directed null vector field normal to ${\cal S}_{u,x}$
is $U:=G^{-1}\partial_x$, and is also tangent to the affinely
parameterised generators of ${\cal N}^+_u$.  The ingoing null normal
on ${\cal S}_{u,x}$ is
\begin{equation}
\label{Xidef}
\Xi=\partial_u -{H\over 2G}\partial_x-bS \partial_y.
\end{equation}
It is normalised to $\Xi^aU_a=-1$.

The field equations we want to solve are the Einstein equations
\begin{equation}
E_{ab}:=R_{ab}-8\pi\nabla_a\psi\nabla_b\psi=0
\end{equation}
and the massless, minimally coupled wave equation
\begin{equation}
\nabla^a\nabla_a\psi=0.
\end{equation}
(We use units where $c=G=1$.) A subset of the Einstein equations, plus
the wave equation, take the form
\begin{eqnarray}
\label{Geqn}
\left(\ln {G\over R_{,x}}\right)_{,x}
&=&S_G[R,f,\psi], \\ 
\label{dbdxeqn}
\left({R^4e^{2S f}b_{,x}\over G}\right)_{,x}&=&S_b[R,f,\psi,G], \\
\label{Reqn}
\left(R\,\Xi R\right)_{,x}&=&S_R[R,f,\psi,G,b], \\
\label{feqn}
\left(R\,\Xi f\right)_{,x}&=&S_f[R,f,\psi,G,b]-(\Xi R)f_{,x},\\
\label{psieqn}
\left(R\,\Xi\psi\right)_{,x}&=&S_\psi[R,f,\psi,G,b]-(\Xi R)\psi_{,x},
\end{eqnarray}
where $\Xi$ is the derivative operator defined in (\ref{Xidef}). We
call these the ``hierarchy equations''. $H$ and $\partial_u$ appear
only in the combination $\Xi$. The right-hand sides $S[f,\dots]$ are
given in full in \cite{axinull_formulation}. They contain the
derivatives $f_{,x}$, $f_{,y}$, $f_{,xy}$ and $f_{,yy}$ (but not
$f_{,xx}$), and similarly for $R$, $G$, $b$ and $\psi$, with
the exception that $\psi_{,xy}$ and $b_{,yy}$ do not appear. 


\subsection{Gauges and solution algorithm}


With $H$ given (for example $H=0$ in double-null gauge), the hierarchy
equations take the form of first-order linear ODEs in $x$ for $b$,
$G$, $\Xi R$, $\Xi f$, and $\Xi\psi$ that can be solved by
integration. With $R$ given (for example $R=x$ in Bondi gauge),
(\ref{Reqn}) is solved for $H$, given $R_{,u}$.  In a third group of
gauges, where $G$ is given (for example $G=1$ in affine gauge),
(\ref{Geqn}) becomes a second-order linear ODE in $x$ for $R$ that is
solved first, the other equations are again solved by integration for
$b$, $\Xi R$ and $\Xi f$, and $(\ref{Reqn})_{,x}$ and
$(\ref{Geqn})_{,u}$ are combined to find an equation that can be
solved for $H$ by integration.

With $f_{,u}$, $R_{,u}$ and $\psi_{,u}$ now known, $f$, $\psi$, and in
double-null gauge also $R$, are now advanced in $u$, and the hierarchy
equations are then solved again. Note that the algorithm is
``maximally constrained'' in the sense that the hypersurface equations
are solved at each time step, as they are from the free initial data.

Consider now the characteristic initial-boundary value problem, with
outgoing null boundary $(u=0,x>0)$ and timelike or null inner boundary
$(x=0,u>0)$. (Geometrically, one can think of the problem on two
intersecting null cones as a pure initial-value problem, but in
Bondi-like coordinates it is more natural to think of data on $u=0$ as
initial data and data on $v=0$ as boundary data, as we solve the
constraints by integration in $v$.)

In double null gauge we specify $R$, $f$ and $\psi$ on
$u=0$, and $\Xi R$, $\Xi f$, $\Xi\psi$, $b$, $b_{,x}$ and $G$ on
$x=0$, or nine functions of two variables. In Bondi gauge we specify
$f$ and $\psi$ on $u=0$, and $\Xi f$, $\Xi\psi$, $b$, $b_{,x}$, $G$
and $H$ on $x=0$, or eight functions of two variables. Finally, in
affine gauge, we specify $f$ and $\psi$ on $u=0$, and $R$, $R_{,x}$,
$\Xi f$, $\Xi\psi$, $b$, $b_{,x}$, $H$ and $H_{,x}$ on $x=0$, or ten
functions of two variables.

In each case, the data on $u=0$ can be specified freely, while the
data on $x=0$ are constrained by some of the remaining Einstein
equations, which do not contain $x$-derivatives. We do not discuss
these constraints here, but evaluate the hyperbolicity of the
evolution equations with the constraints relaxed.


\subsection{Definition of strong hyperbolicity}


As already noted, a necessary condition for well-posedness of a
nonlinear PDE is well-posedness of its linearisation, in the principal
part, frozen coefficient approximation.  We denote the field equations
linearised around a background solution $\phi$ by 
\begin{equation}
L({\bf x},\phi,\nabla)\delta\phi=0,
\end{equation}
We denote the principal part of $L$ by $L^p$. The principal symbol is
the matrix-valued function $L^p({\bf x},\phi,i{\bf k})$ of the wave number
covector ${\bf k}$.  For a quasilinear system (such as the Einstein
equations), the principal symbol of the linearisation is the same as
the principal part of the full equations.

We can use the principal symbol to find plane-wave solutions to the
linearised field equations in the frozen coefficient approximation:
substituting $\delta\phi({\bf x})=e^{i{\bf k}\cdot{\bf x}}\widehat{\delta\phi}({\bf k})$
into $L(\nabla)\delta\phi({\bf x})=0$ gives
$L(i{\bf k})\widehat{\delta\phi}({\bf k})=0$, which is now a system of linear
algebraic equations. Nontrivial solutions exist only for ${\bf k}$ such
that $\det L(i{\bf k})=0$. These ${\bf k}$ are called characteristic covectors,
and vectors $\widehat{\delta\phi({\bf k})}$ in the corresponding null space
of $L(i{\bf k})$ are called characteristic variables. The linearised
problem with frozen coefficients is well-posed in $L^2$ if the
plane-wave solutions with real ${\bf k}$ are complete in the sense that we
can map them smoothly and one-to-one to the boundary and initial data.

Consider now the Cauchy problem for a system of linear PDEs with
constant coefficients. Initial data are imposed on a spacelike surface
$t=0$, with normal covector ${\bf n}:=dt$, and we assume that the initial
data fall off as $|{\bf x}|\to 0$, so that we can Fourier-transform the
initial data and the solution in ${\bf x}$. Then a sufficient criterion for
well-posedness is given by strong hyperbolicity. We offer the
following definitions:

A system of PDE is defined to be {\bf strictly hyperbolic} in the time
direction ${\bf n}$, in a neighbourhood of the solution $\phi$, if, first,
$\det L^p(\phi,i{\bf n})\ne 0$, and second, all roots $\omega$ of $\det
L^p(\phi,i\omega{\bf n}+i{\bar{\bf k}})=0$ are real and distinct, for all real ${\bar{\bf k}}$
that are not zero or proportional to ${\bf n}$ (see, for example,
\cite{RenardyRogers}). (We shall use the notation ${\bf k}$ for arbitrary
covectors, ${\bar{\bf k}}$ for what intuitively are vectors in ``space'', and
${\bf k}_\omega:=\omega{\bf n}+{\bar{\bf k}}$.)

If the roots are not all distinct, then the system is {\bf strongly
  hyperbolic} if the null spaces of $L^p(\phi,i\omega{\bf
  n}+i{\bar{\bf k}})$ have dimension corresponding to the multiplicity
of each root $\omega$, and these null spaces depend smoothly on
$\omega$. This generalises the textbook definition for first-order
systems (see, for example, \cite{GKO}).

If the roots are all real, but the system is not strongly hyperbolic,
it is called {\bf weakly hyperbolic}.

Strict hyperbolicity implies strong hyperbolicity, and for the
linearised equations with frozen coefficients both mean that after
Fourier-transforming in ${\bf x}$, initial data on a surface with
normal covector ${\bf n}$ can be decomposed into plane waves
$\exp(i{\bf k}\cdot{\bf x})\delta\phi({\bf k})$, with $\delta\phi({\bf
  k})$ in the null spaces mentioned above, each of which propagates
with a speed $\omega$ in the direction $-{\bf k}$. This in turn
implies local well-posedness of the pure initial-value problem.

If we reduce a strongly hyperbolic system of arbitrary higher order to
first order by introducing suitable derivatives of the original
variables as reduction variables, the general solution translated into
these variables still consists of purely oscillating plane
waves. Hence any such first-order reduction is strongly
hyperbolic. However, the converse is not true: a higher order system
that is only weakly hyperbolic according to our definition may or may
not admit a strongly hyperbolic first-order reduction.


\subsection{The principal symbol}
\label{section:principalsymbol}


To fix notation, we write our variables in the order
$\phi^\dagger:=(G,b,R,f,H,\psi)$, the coordinates in the order
${\bf x}:=(u,x,y)$, so that $\nabla:=(\partial_u,\partial_x,\partial_y)$,
and we use the notation ${\bf k}:=(\mu,\xi,\eta)$ for the wave number
covector.

For our field equations, the principal terms clearly include the
second derivatives of $G$, $b$, $R$, $f$ and $\psi$. The only
derivative of $H$ that appears is $H_{,x}$, and this should therefore
be considered principal. Finally, because we want to think of
(\ref{Geqn}) as determining $G$, we include the first derivative
$G_{,x}$ as a principal term in this equation, but not in the other
equations, which also contain second derivatives of $G$. The equations
are quasilinear in this sense. 

Note that until we have fixed a specific null gauge, the principal
symbol will have five rows, corresponding to the linearisations of the
field equations (\ref{Geqn}-\ref{psieqn}), but six columns,
corresponding to six metric perturbations. The preliminary principal
symbol and the corresponding vector of perturbation variables are
\begin{widetext}
\begin{equation}
\label{5by6}
L^p({\bf x},\phi,i{\bf k})=\left(
\begin{array}{cccccc}
{i\over G}\xi & 0 & {1\over R_{,x}}\xi^2 & 0 & 0 & 0\\
-{R^2\over G}\eta\,\xi & -{e^{2S f}R^4\over G}\xi^2 & -
2R\,\eta\,\xi & 2R^2S\, \eta\,\xi & 0  & 0\\
{R^2 \over 4}Y & {R^2S\over 4}\eta\,\xi &
X+{GR \over 2}Y &
-{GR^2S \over 2}Y & -i{RR_{,x}\over 2G}\xi & 0 \\
{R\over 4S}Y & {R\over 4}\eta\,\xi & 0 &
X & -i{R f_{,x}\over 2G}\xi  & 0\\
0 & 0 & 0 & 0 & -i{R \psi_{,x}\over 2G}\xi & 
X+{GR \over 2}Y \\
\end{array}
\right),
\qquad
\delta\phi({\bf x}):=\left(\begin{array}{c}
\delta G \\ \delta b \\ \delta R
  \\ \delta f \\ \delta H \\ \delta\psi
\end{array}\right).
\end{equation}
\end{widetext}

We have defined the shorthands
\begin{eqnarray}
X&:=&R\,\xi\left(-\mu+b S \,\eta+{H\over 2G}\,\xi\right), \\
Y&:=&{e^{-2S f}S\over R^2}\eta^2.
\end{eqnarray}
$X$ is the principal symbol of the derivative operator
$\partial_xR\Xi$ that appears on the left-hand side of each of the
hierarchy equations.  $Y$ is the symbol of the principal angular
derivative $-g^{yy}\partial_y\partial_y$.

Each of the three classes of Bondi-like gauges that we have already
discussed simply eliminates one of the columns of (\ref{5by6}), giving
us a square $5\times 5$ symbol: this is the $\delta H$ column in
generalised double-null gauges, where $H$ is given and $\delta H=0$;
the $\delta R$ column in Bondi and related gauges, where $R$ is given
and $\delta R=0$; and the $\delta G$ column in affine and related
gauges, where $G$ is given and $\delta G=0$. We do not
write the $5\times 5$ principal symbols for these gauges out
explicitly, as they can be read off trivially from (\ref{5by6}).

In any null gauge, the ``time'' direction ${\bf n}=du=(1,0,0)$, that is, using
$u$ as the time coordinate, fails the first criterion for strong
hyperbolicity. That is of course expected, as the surfaces of constant
$u$ are characteristic. Instead we follow \cite{GHZ2020} and analyse
hyperbolicity on spacelike time slices with normal covector
\begin{equation}
\label{ndef}
{\bf n}:=\left(1,{1\over A},0\right), 
\end{equation}
with $A>0$. For now we leave $A$ and ${\bar{\bf k}}=(\mu,\xi,\eta)$ general, so
that
\begin{equation}
{\bf k}_\omega=\left(\omega+\mu,{\omega\over A}+\xi,\eta\right). 
\end{equation}


\subsection{Double-null and related gauges}
\label{section:doublenull}


In double-null gauge $H=0$ and $\delta H=0$. In related gauges
\cite{axinull_formulation}, $\delta H$ is given in terms of $\delta
G$, and so the resulting $\delta H_{,x}$ is not principal in the
equations where it appears, because these contain second derivatives
of $G$. In either case we can delete the $\delta H$ column in the
temporary principal symbol to obtain the $5\times 5$ principal symbol
for these gauges, and we remover $\delta H$ from $\delta\phi$. 

The first row of $L^p(\phi,i{\bf k})$ now contains first and second order
in ${\bf k}$ terms, and the other rows are homogeneous in ${\bf k}$ of second
order. However, the cofactor of the entry $\xi^2/R_{,x}$ in the
first row, third column, of $L^p(\phi,i{\bf k})$ is zero, and so $\det
L^p(\phi,i{\bf k})$ is homogeneous of order nine, rather than ten, in ${\bf k}$.

The nine roots $\omega$ of $L^p(\phi,i\omega{\bf n}+i{\bf k})=0 $ are
\begin{equation}
\omega_0:=-A\xi
\end{equation}
with multiplicity four,
\begin{equation}\omega_\pm:=\omega_c\pm\Delta\omega,
\end{equation}
each with multiplicity two, and
\begin{equation}
\omega_2:=2\omega_c+A\xi=-2Z\bar\mu-A\xi, 
\end{equation}
with multiplicity one. Here
we have defined the shorthands
\begin{eqnarray}
\omega_c&:=&-Z\bar\mu-A\xi, \\
\label{omegadeltadef}
\Delta\omega&:=&\sqrt{Z^2\bar\mu^2
+AGZ\,Y}, \\
\bar\mu&:=&\mu-A\xi-bS\eta, \\
Z&:=&{AG\over 2AG-H}.
\end{eqnarray}
The roots $\omega_\pm$ with multiplicity two have corresponding null
spaces of dimension two, and the null space of
$\omega_2$ has dimension one, but the null space of $\omega_0$ has
only dimension two, not four,  so two fewer than the multiplicity of the root.

Attention must also be given to special directions in which roots
$\omega$ merge. In the special direction given by $\bar\mu=0$, the
roots $\omega_2=\omega_0$ have merged and $\omega_0$ is now a
quintuple root, but the corresponding null space is still only
two-dimensional, so (exceptionally) three fewer than the multiplicity
of the root.

The characteristic covectors ${\bf k}_\omega$ themselves are independent of
the parameter $A$ of the time slicing, as long as $A>0$, and to find
all of them it is sufficiently general to give ${\bar{\bf k}}$ only two
algebraically independent components. In particular, ${\bf k}_\pm$ and
${\bf k}_2$ can be written as
\begin{eqnarray}
\label{kpmdef}
{\bf k}_\pm&=&\left({GY\over 2\tilde\xi}+{H\over
  2G}\tilde\xi+bS\eta,\tilde\xi,\eta\right), \\ {\bf k}_2&=&\left({H\over
  2G}\tilde\xi+bS\eta,\tilde\xi,\eta\right),
\end{eqnarray}
where now $\tilde\xi$ and $\eta$ are arbitrary, except that
$\tilde\xi>0$ for ${\bf k}_+$ and $\tilde\xi<0$ for ${\bf k}_-$. Similarly,
${\bf k}_0$ can be written as
\begin{equation}
{\bf k}_0=(\tilde\mu,0,\eta), 
\end{equation}
where now $\tilde\mu$ and $\eta$ are arbitrary. We see that
${\bf k}_0$ parameterises plane waves that do not depend on $x$. The
${\bf k}_\pm$ are null covectors, while ${\bf k}_0$ and ${\bf k}_2$ are
spacelike. They obey $U\cdot {\bf k}_0=0$ and $\Xi\cdot
  {\bf k}_2=0$.


\subsection{Bondi and related gauges}


In Bondi and related gauges, $R$ is given and $\delta R=0$, and so we
eliminate the column corresponding to it. We now have eight roots
$\omega$: $\omega_0$ with multiplicity four and null space of
dimension two, so two fewer than the multiplicity of the root, and
$\omega_\pm$ with multiplicity two and the same null spaces of
dimension two as in the double null case. There is no problem of
merging roots.


\subsection{Affine and related gauges}
\label{section:affine}


In affine gauge, $G$ is given and $\delta G=0$, and so we eliminate
the column corresponding to it. With $G=G(u)$,
  (\ref{Geqn}) and (\ref{Reqn}) take the form
\begin{eqnarray}
\label{Geqnaffine}
R_{,xx}&=&(...)R, \\
\label{Reqnaffine}
(RR_{,u})_{,x}+{1\over 2G}(RR_{,x}H)_{,x}&=&(...),
\end{eqnarray}
where the dots stand for terms already known at the point where the
equations are solved. (\ref{Geqnaffine}) is then solved as a
second-order ODE in $x$ for $R$.  Starting from
$(\ref{Reqnaffine})_{,x}$, we use $(\ref{Geqnaffine})_{,u}$ to
eliminate $R_{,uxx}$ and $(\ref{Reqnaffine})$ to eliminate $R_{,ux}$,
and obtain an equation of the form $H_{,xx}=(...)$. 

To reflect this, in affine gauge we multiply the third row of
(\ref{5by6}), which comes from (\ref{Reqn}), by $i\xi$. This will
obviously multiply $\det L_p(i{\bf k})$ by a factor of $i\xi$, and hence
will gives us an extra root $\omega_0$, without changing the dimension
of the corresponding null space.

We now have ten roots $\omega$: $\omega_0$ with multiplicity six and
null space of dimension three, so three fewer than the multiplicity of
the root and $\omega_\pm$ with multiplicity two and the same null
spaces of dimension two as in the double null case. There is no
problem of merging roots.


\subsection{Linearisation about Minkowski spacetime and frozen
  coefficient approximation}


The principal symbol, and hence the characteristic covectors and
variables are simpler in the linearisation about Minkowski, but the
multiplicities of the roots and dimensions of the corresponding null
spaces are unchanged, so we now restrict to this case.

Linearising about Minkowski spacetime, so that $b=f=\psi=0$, $G=H=1$
and $R=x$ in the background, we have the $5\times 6$ principal symbol
\begin{widetext}
\begin{equation}
\label{5by6linMink}
L^p_\text{Mink}({\bf x},i{\bf k})=\left(
\begin{array}{cccccc}
i\xi & 0 & \xi^2 & 0 & 0 & 0 \\
-r^2\eta\xi & -r^4\xi^2 & -2r\eta\xi & 2r^2S\eta\xi & 0 & 0\\
{S\over 4}\eta^2 & {r^2S\over 4}\eta\xi & 
{S\over 2r}\eta^2-r\mu\xi+{r\over 2}\xi^2  &
-{S^2 \over 2}\eta^2 & -i{r\over 2}\xi & 0 \\
{1\over 4r}\eta^2 & {r\over 4}\eta\xi & 0 &
-r\mu\xi+{r\over2}\xi^2 & 0 & 0 \\
0 & 0 & 0 & 0 & 0 & {S\over 2r}\eta^2-r\mu\xi+{r\over 2}\xi^2 \\
\end{array}\right).
\end{equation}
In the frozen coefficient approximation, we denote the frozen value of
$R$ in the background by $r$, but the radial argument of the linear
perturbations by $x$. 

We can simplify (\ref{5by6linMink}) further, as follows. We define the
coordinate ${\bar y}$ and corresponding wave number $k$ as
\begin{equation}
\label{kdef}
k:={\sqrt{S}\over r}\eta, \quad {\bar y}:={r\over\sqrt{S}}y,
\quad\Rightarrow\quad
k{\bar y}=\eta y, \quad \partial_y={r\over\sqrt{S}}\partial_{\bar y}.
\end{equation}
Note that in the frozen coefficient approximation both $y$ and ${\bar
  y}$ are considered to have infinite range, while $S$ appears only as
a frozen coefficient, ignoring that $S=1-y^2=\sin^2\theta$. Similarly,
$r$ stands for a frozen coefficient, while $x$ is considered to have
infinite range. We multiply the rows of (\ref{5by6linMink}) by the
constants $(1,\sqrt{S},4,4S,2)$ respectively, and renormalise the
perturbations by constants as
\begin{equation}
\overline{\delta\phi}^\dagger:=(\delta G,(r\sqrt{S})\delta b,(2/r) \delta
  R,(2 S)\delta f,(2/r)\delta H,\delta\psi) 
:=(\delta G,\overline{\delta b},
\overline{\delta R},\overline{\delta f},\overline{\delta H},
\delta\psi),
\end{equation}
which corresponds to multiplying the columns of (\ref{5by6linMink}) by
constants. With this notation, the preliminary $5\times 6$ principal
symbol of the linearisation about Minkowski space takes the form
\begin{equation}
\label{5by6linMinkiren}
L^p_\text{Mink,frozen}(i{\bf k})=\left(
\begin{array}{cccccc}
i\xi & 0 & {r\over 2}\xi^2 & 0 & 0 & 0 \\
-k\xi & -\xi^2 & - k\xi & k\xi & 0 & 0\\
k^2 & k\xi & k^2-2\mu\xi+\xi^2 & -k^2 & -i\xi & 0 \\
k^2 & k\xi & 0 & -2\mu\xi+\xi^2 & 0 & 0 \\
0 & 0 & 0 & 0 & 0 & k^2-2\mu\xi+\xi^2 \\
\end{array}\right).
\end{equation}
\end{widetext}
We have been able to absorb all but one of the background-dependent
coefficients into our redefinitions.

We also simplify the calculation of characteristic covectors by
restricting our ansatz to ${\bf n}:=(1,1,0)$ and $\bar k=(0,\xi,k)$, so that
\begin{equation} 
{\bf k}_\omega=(\omega,\omega+\xi,k).
\end{equation}
In all three classes of Bondi-like gauges, $\det L^p(i{\bf k}_\omega)=0$ has roots
\begin{equation}
\omega_\pm=\pm\sqrt{\xi^2+k^2}, 
\end{equation}
corresponding to characteristic covectors
\begin{equation}
\label{kpmzetak}
{\bf k}_\pm=\left(\pm\sqrt{\xi^2+k^2},\xi\pm\sqrt{\xi^2+k^2},k\right), 
\end{equation}
with multiplicity two and null spaces of
dimension two. In addition, in double null gauge we have
$\omega_0=-\xi$, or
\begin{equation}
{\bf k}_0=\left(-\xi,0,k\right), 
\end{equation}
with multiplicity four but null space of dimension two,
and $\omega_2=\xi$, or
\begin{equation}
{\bf k}_2=\left(\xi,2\xi,k\right), 
\end{equation} with multiplicity one.
In Bondi gauge we have $\omega_0$ with multiplicity four but null
space of dimension two, and in affine gauge $\omega_0$ with multiplicity
six but null space of dimension four. So in each gauge the
obstruction to strong hyperbolicity is that the multiplicity of the
characteristic vector ${\bf k}_0$ is two more than the dimension of the
corresponding null space.

In the special case $\xi=0$, we have
\begin{equation}
{\bf k}_\pm=(\pm k,\pm k,k), \qquad {\bf k}_2={\bf k}_0=(0,0,k),
\end{equation}
and in the special case $k=0$, we have
\begin{equation}
\label{specialcasek=0}
{\bf k}_+={\bf k}_2=(\xi,2\xi,0), \qquad {\bf k}_-={\bf k}_0=(-\xi,0,0).
\end{equation}
For $\xi=k=0$, there are no characteristic covectors.


\subsection{A toy model for the obstruction to strong hyperbolicity}


\begin{table*}
\setlength{\tabcolsep}{8pt} 
\renewcommand{\arraystretch}{1.} 
\begin{tabular}{c|cc|c|c|c}
gauge & init. data  & boundary data &  characterist. covecs. &  free functions & count \\
\hline
double null & $R,f$ & $b,b_{,x},G,\Xi R,\Xi f$ & ${\bf k}_\pm,{\bf k}_2,{\bf k}_0\times4$ 
& $\hat c_\pm,\hat c_2,\overline{\delta G}_0,\overline{\delta
  b}_0,\overline{\delta b}_1,\overline{\delta f}_0 $ & 7 \\
Bondi & $f$ & $b,b_{,x},G,\Xi f,H$ & ${\bf k}_\pm,{\bf k}_0\times 4$ 
& $\hat c_\pm,\overline{\delta G}_0,\overline{\delta
  b}_0,\overline{\delta f}_0,\overline{\delta H}_0$ & 6\\
affine & $f$ &$b,b_{,x},R,R_{,x},\Xi f,H,H_{,x}$ & 
${\bf k}_\pm,{\bf k}_0\times 6$ 
& $\hat c_\pm,\overline{\delta b}_0,\overline{\delta
  R}_0,\overline{\delta R}_1,\overline{\delta f}_0,\overline{\delta
  H}_0,\overline{\delta H}_1$ & 8\\
\hline
scalar field & $\psi$ & $\Xi\psi$ & ${\bf k}_\pm$  & $\hat \psi_\pm$
& 2 \\
\end{tabular}
\caption{Function counting for the general solution (in the frozen
  coefficient approximation) versus free initial and boundary data.
  Here all barred functions are arbitrary functions of $(u,y)$, while
  all hatted functions are arbitrary functions of $(\xi,\eta)$. The
  mixed physical space/Fourier space notation has been chosen simply
  for ease of function counting, but see
  (\ref{deltapsiFourier},\ref{PsiFourier},\ref{Psi2Fourier}) for how
  to translate everything into real space.  The massless scalar matter
  field decouples from the metric perturbations in the linearisation
  about Minkowski, in any gauge, and is therefore listed
  separately. Initial data are imposed on $u=0$ and boundary data are
  imposed on $v=0$ or $r=r_0$.}
\label{table:multiplicities}
\end{table*}

If a weakly hyperbolic system of homogeneous linear PDEs
with constant coefficients fails to be strongly hyperbolic because
there are not enough plane-wave solutions, the missing solutions must
be a polynomial times a plane wave. (The reader is reminded of a
well-known textbook example in Appendix~\ref{appendix:toysystem}.)
 
We now use the principal part, frozen coefficient approximation as a
toy model for the full linearised equations. The fact that it has
constant coefficients allows us to find the general solution in closed
form, and in particular the polynomial solutions that obstruct strong
hyperbolicity.

In the linearisation about Minkowski spacetime the scalar wave
equation decouples from the metric perturbation in any choice of
Bondi-like null gauge, and so we can consider it separately. In the
frozen coefficient approximation, it takes the form
\begin{equation}
\label{Minkowskiwaveeqn}
-2\delta\psi_{,ux}+\delta\psi_{,xx}+\delta\psi_{,{\bar y}{\bar y}}=0.
\end{equation}
Note this is the actual wave equation on 2+1-dimensional Minkowski
spacetime with metric
\begin{equation}
\label{2+1Mink}
ds^2=-2\,du\,dx-du^2+dx^2+d{\bar y}^2.
\end{equation}
A plane-wave ansatz gives the solution
\begin{equation}
\label{deltapsiFourier}
\delta\psi(u,x,{\bar y}) =\sum_\pm \int \int
\widehat{\delta\psi}_\pm(\xi,k) e^{i{\bf k}_\pm\cdot{\bf x}}\,d\xi\,dk.
\end{equation}
The functions $\widehat{\delta\psi}_\pm(\xi,k)$ map one-to-one to, for
example, characteristic data in $L^2$ on null cones $u=0$ and
$v:=u+2x=0$, or to Cauchy data in $L^2$ on $t:=u+x=0$.  Hence this
solution is complete by function counting, with no polynomial
solutions required.

We now derive the general solution for the metric perturbations. We
restrict to vacuum, $\delta\psi=0$, without loss of generality.
As the missing plane-wave solutions are for ${\bf k}_0=(-\xi,0,k)$, which
parameterises functions that are independent of $x$, we expect the
missing solutions to be polynomial in $x$. The physical gravitational
waves can be expressed in terms of derivatives of a solution $\Psi$ of
the {\em scalar} wave equation \cite{GomezPapadopoulosWinicour1994,axinull_formulation}, which can be parameterised as
\begin{equation}
\label{PsiFourier}
\Psi(u,x,{\bar y}) =\sum_\pm \int \int
\hat c_\pm(\xi,k) e^{i{\bf k}_\pm\cdot{\bf x}}\,d\xi\,dk,
\end{equation}
in complete parallel to (\ref{deltapsiFourier}). In double-null gauge,
we also introduce the shorthand
\begin{equation}
\label{Psi2Fourier}
\Psi_2(u,x,{\bar y}) = \int \int
\hat c_2(\xi,k) e^{i{\bf k}_2\cdot{\bf x}}\,d\xi\,dk.
\end{equation}
for the general solution in $L^2$ of the advection equation
\begin{equation}
2\Psi_{2,u}-\Psi_{2,x}=0
\end{equation}
along ingoing null cones.

In {\bf double null gauge} and its generalisations, we set
$\overline{\delta H}=0$. The equations in the principal part, frozen coefficient
approximation can be read off from (\ref{5by6linMinkiren}) with the
$\widehat{\delta H}$ column deleted. They are
\begin{eqnarray}
\label{myGeqndn}
\delta G_{,x}-{r\over 2}\,\overline{\delta R}_{,xx}&=&0, \\
\label{mybeqndn}
\delta G_{,x{\bar y}}+\overline{\delta b}_{,xx}
+\overline{\delta R}_{,x{\bar y}}
-\overline{\delta f}_{,x{\bar y}}&=&0, \\
\label{myHeqndn}
-\delta G_{,{\bar y}{\bar y}}
-\overline{\delta b}_{,x{\bar y}}
+2\overline{\delta R}_{,ux}-\overline{\delta R}_{,xx}
-\overline{\delta R}_{,{\bar y}{\bar y}} 
\nonumber \\ 
+\overline{\delta f}_{,{\bar y}{\bar y}}&=&0, \\
\label{myfeqndn}
-\delta G_{,{\bar y}{\bar y}}-\overline{\delta b}_{,x{\bar y}}
+2\overline{\delta f}_{,ux}\ -\overline{\delta f}_{,xx}&=&0.
\end{eqnarray}
The general solution is
\begin{eqnarray}
\delta G  &=& \delta G_{0,u} + {r\over 2}\Psi_{2,x},   \\
\overline{\delta b}  &=& \overline{\delta b}_0 
+ \overline{\delta b}_{1,u} x -  {r\over 2} \Psi_{2,{\bar y}} +\Psi_{,{\bar y}}, \\
\overline{\delta R} &=& \overline{\delta f}_0 
+ {1\over 2}\left(\overline{\delta b}_{1,{\bar y}} + \delta G_{0,{\bar y}{\bar y}}\right)x 
+ \Psi_2, \\
\overline{\delta f}  &=& \overline{\delta f}_0 
+ {1\over 2}\left(\overline{\delta b}_{1,{\bar y}} + \delta G_{0,{\bar y}{\bar y}}\right)x 
+\Psi_2  +\Psi_{,x}.
\end{eqnarray}
The five free functions (of two variables) $\hat c_\pm$, $\hat c_2$,
$\overline{\delta b}_0$, $\overline{\delta f}_0$ parameterise plane
waves, while the free functions $\delta G_0$ and $\overline{\delta
  b}_1$ parameterise linear-in-$x$ solutions. These seven free
functions correspond to seven characteristic covectors: ${\bf k}_\pm$ and
${\bf k}_2$, all with multiplicity one, and ${\bf k}_0$ with multiplicity
four. They also correspond to the freedom to set five functions
$(b,b_{,x},G,\Xi R,\Xi f)$ of $(u,{\bar y})$ at $x=0$ and two functions
$(R,f)$ of $(x,{\bar y})$ at $u=0$. Hence our solution is complete by
function counting.

The linear-in-$x$ solutions are problematic not because they grow (and
so are not in $L^2$), but because they grow arbitrarily rapidly in $x$
for boundary data at $x=0$ that oscillate arbitrarily rapidly in $y$.

In {\bf Bondi gauge} and its generalisations, we set $\overline{\delta
  R}=0$. The equations in the principal part, frozen coefficient
approximation can be read off from (\ref{5by6linMinkiren}) with the
$\widehat{\delta R}$ column deleted. They are
\begin{eqnarray}
\label{myGeqnB}
\delta G_{,x}&=&0, \\
\label{mybeqnB}
\delta G_{,x{\bar y}}+\overline{\delta b}_{,xx}-\overline{\delta
  f}_{,x{\bar y}}&=&0, \\
\label{myHeqnB}
-\delta G_{,{\bar y}{\bar y}}-\overline{\delta
  b}_{,x{\bar y}}+\overline{\delta f}_{,{\bar y}{\bar
    y}}-\overline{\delta H}_{,x}&=&0, \\
\label{myfeqnB}
-\delta G_{,{\bar y}{\bar y}}-\overline{\delta b}_{,x{\bar y}}+2\overline{\delta
  f}_{,ux}\ -\overline{\delta
  f}_{,xx}&=&0.
\end{eqnarray}
The general solution is
\begin{eqnarray}
\label{Bondifirst}
\delta G  &=& \delta G_0,   \\
\overline{\delta b}  &=& \overline{\delta b}_0 
-\delta G_{0,{\bar y}} x + \Psi_{,{\bar y}}, \\
\overline{\delta f}  &=& \overline{\delta f}_0  
+\Psi_{,x}, \\
\overline{\delta H}  &=& \overline{\delta H}_0 
+\overline{\delta f}_{0,{\bar y}{\bar y}} x.
\label{Bondilast}
\end{eqnarray}
The four free functions $\hat c_\pm$, $\overline{\delta b}_0$ and
$\overline{\delta H}_0$ parameterise plane waves, while
$\delta G_0$ and $\overline{\delta f}_0$ parameterise
linear-in-$x$ solutions. These six free functions correspond to six
characteristic covectors: ${\bf k}_\pm$ with multiplicity one and ${\bf k}_0$
with multiplicity four. They also correspond to the freedom to set
five functions $(b,b_{,x},G,\Xi f,H)$ at $x=0$ and one function $f$ at
$u=0$. Hence our solution is complete by function counting.  

Finally, in {\bf affine gauge} and its generalisations, we set
$\delta G=0$. The equations in the principal part, frozen coefficient
approximation can be read off from (\ref{5by6linMinkiren}) with the
$\widehat{\delta G}$ column deleted. They are
\begin{eqnarray}
\label{myGeqnaf}
\overline{\delta R}_{,xx}&=&0, \\
\label{mybeqnaf}
\overline{\delta b}_{,xx}
+\overline{\delta R}_{,x{\bar y}}
-\overline{\delta f}_{,x{\bar y}}&=&0, \\
-\overline{\delta b}_{,x{\bar y}}
+2\overline{\delta R}_{,ux}-\overline{\delta R}_{,xx}
-\overline{\delta R}_{,{\bar y}{\bar y}} 
\nonumber \\ 
\label{myHeqnaftmp}
+\overline{\delta f}_{,{\bar y}{\bar y}}-\overline{\delta H}_{,x}&=&0, \\
\label{myfeqnaf}
-\overline{\delta b}_{,x{\bar y}}
+2\overline{\delta f}_{,ux}\ -\overline{\delta f}_{,xx}&=&0.
\end{eqnarray}
Following what is done in the full nonlinear equations, we take an
$x$-derivative of (\ref{myHeqnaftmp}), and use derivatives of the
other equations to simplify it. In the principal part, frozen coefficient
approximation the result is simply
\begin{equation}
\label{myHeqnaf}
\overline{\delta H}_{,xx}=0.
\end{equation}

The general solution of (\ref{myGeqnaf}-\ref{myfeqnaf},\ref{myHeqnaf})
is
\begin{eqnarray}
\overline{\delta b}  &=& \overline{\delta b}_0 
+2{\delta R}_{1,u} x + \Psi_{,{\bar y}}, \\
\overline{\delta R}  &=& \overline{\delta R}_0 + \overline{\delta R}_{1,{\bar y}} x,  \\
\overline{\delta f}  &=& \overline{\delta f}_0 + \overline{\delta R}_{1,{\bar y}} x
+\Psi_{,x}, \\
\overline{\delta H}  &=& \overline{\delta H}_0 
+\overline{\delta H}_1 x.
\end{eqnarray}
The six free functions $\hat c_\pm$, $\overline{\delta b}_0$,
$\overline{\delta H}_0$, $\overline{\delta R}_0$ and $\overline{\delta
  f}_0$ parameterise plane waves, while $\overline{\delta R}_1$ and
$\overline{\delta H}_1$ parameterise linear-in-$x$ solutions. These
eight free constants correspond to eight characteristic covectors:
${\bf k}_\pm$ with multiplicity one and ${\bf k}_0$ with multiplicity six. They
also correspond to the freedom to set seven functions
$(b,b_{,x},G,R,R_{,x},\Xi f,H,H_{,x})$ at $x=0$ and one function $f$
at $u=0$. Hence our solution is once again complete by function
counting.

If we solved (\ref{myHeqnaftmp}) instead of (\ref{myHeqnaf}) then
$\overline{\delta H}_1$ would not be free, but given by
$\overline{\delta H}_1=\overline{\delta f}_{0,{\bar y}{\bar y}}-\overline{\delta
  R}_{0,{\bar y}{\bar y}}$. However, this is a feature of the toy
model only, where are able to solve in closed form for
$\overline{\delta R}$.

Table~\ref{table:multiplicities} gives an overview of free null data
and normal data, the characteristic covectors, and the free functions
in our explicit solution (in the frozen-coefficient approximation, in
the linearisation about Minkowski). In each case, if we add
$\overline{\delta\psi}$ back in, the number of free function, matching
the free data, becomes 9, 8 and 10 respectively, as discussed above in
Sec.~\ref{section:doublenull}-\ref{section:affine}. In all gauges, the
boundary data at $x=0$ are subject to constraints, which we do not
impose here.


\section{Symmetric hyperbolic first-order
  reduction of the linearised Einstein equations in Bondi gauge}
\label{section:Frittelli}



\subsection{Balance laws from symmetric hyperbolic first-order systems}
\label{section:balancelaw}


This subsection reviews relevant parts of \cite{Frittelli2005b} for
completeness and to establish notation. A system of linear first-order PDEs
\begin{equation}
\label{phisystem}
C^\mu\phi_{,\mu}=D\phi,
\end{equation}
where $\phi\in{\Bbb R}^N$ is a vector of dependent variables, $C^\mu$ and
$D$ are real $N\times N$ matrices that depend smoothly on
$x^\mu\in{\Bbb R}^n$, and the $C^\mu$ are symmetric, implies the balance law
\begin{eqnarray}
\label{balancelaw}
{j^\mu}_{,\mu}&=&S, \\
j^\mu&:=&\phi^\dagger C^\mu\phi, \\
S&:=&\phi^\dagger{\cal D}\phi, \\
{\cal D}&:=&D+D^\dagger+{C^\mu}_{,\mu}.
\end{eqnarray}
The matrix ${\cal D}$ is real and symmetric, and so can be
diagonalised with real eigenvalues.

If furthermore $C^\mu k_\mu$ depends smoothly on the covector
$k_\mu\ne 0$, Gauss' law gives
\begin{equation}
\label{integratedbalancelaw}
\int_{\partial V}j^\mu d\Sigma_\mu=\int_V S\,dV
\end{equation}
for any ``control'' spacetime volume $V$ with boundary $\partial V$.

If furthermore there exists a time coordinate $t$ (a smooth function
with $(dt)_\mu$ everywhere future pointing), such that
\begin{equation}
j^t:=j^\mu (dt)_\mu
\end{equation}
is positive definite in $\phi$, the system is called symmetric
hyperbolic with respect to $t$.
 
We now consider two elementary generalisations. If we redefine the
dependent variables as
\begin{equation}
\label{phitildedef}
\phi:=A\tilde\phi,
\end{equation}
where $A$ is invertible and depends smoothly on the
$x^\mu$, (\ref{phisystem}) becomes
\begin{equation}
\tilde C^\mu \tilde\phi_{,\mu}=\tilde D\tilde\phi,
\end{equation}
where
\begin{eqnarray}
\tilde C^\mu&:=&A^\dagger C^\mu A, \\
\label{tildeDdef}
\tilde D&:=&A^\dagger(DA-C^\mu A_{,\mu}).
\end{eqnarray}
The resulting balance law is the same as before, namely
\begin{eqnarray}
\tilde j^\mu&:=&\tilde\phi^\dagger\tilde C^\mu\tilde\phi=j^\mu, \\
\tilde S&:=&\tilde\phi^\dagger\tilde{\cal D}\tilde\phi=S, \\
\label{tildecalDdef}
\tilde {\cal D}&:=&A^\dagger{\cal D}A,
\end{eqnarray}
In particular, the $A_{,\mu}$ term in $\tilde D$ in (\ref{tildeDdef})
cancels out of $\tilde{\cal D}$ in (\ref{tildecalDdef}). However, we
can change the eigenvalues of $\tilde{\cal D}$ and $\tilde C^\mu$ by
choosing an invertible but non-orthogonal $A$. To see this, note that
because ${\cal D}$ is symmetric, we have
\begin{equation}
{\cal D}=R\Lambda R^{\dagger}, \qquad R^\dagger R=I,
\end{equation}
so
\begin{equation}
\tilde{\cal D}=A^\dagger R\Lambda R^{\dagger}A,
\end{equation}
Hence $\tilde{\cal D}$ has the same eigenvalues as ${\cal D}$ if and
only if $A$ is orthogonal.  The same is true for the matrices
the $\tilde C^\mu$. 

As a second modification, we consider a change of integration weight
$dV$ in (\ref{integratedbalancelaw}). The balance law
(\ref{balancelaw}) is equivalent to
\begin{equation}
\label{omegaconservationlaw}
(\omega{j^\mu})_{,\mu}=\omega S_\omega
\end{equation}
with
\begin{equation}
\label{Somegadef}
S_\omega:=S+{\omega_{,\mu}\over\omega}j^\mu
=\phi^\dagger\left({\cal D}+{\omega_{,\mu}\over\omega}C^\mu\right)\phi,
\end{equation}
and so we have
\begin{equation}
\label{integratedbalancelaw}
\int_{\partial V}j^\mu \,\omega d\Sigma_\mu=\int_V S_\omega\,\omega dV
\end{equation}
for any smooth function $\omega>0$.


\subsection{Estimate on an arbitrary control volume}
\label{section:generalestimates}


We now review how a symmetric hyperbolic system of
homogeneous first-order PDEs gives rises to ``energy'' estimates that
demonstrate well-posedness in the corresponding ``energy norm''.  We
follow the basic idea of \cite{Frittelli2005b}, but give an
alternative derivation.

Let $n_\mu\ne 0$ denote an outward-pointing covector field on
$\partial V$. As the $C^\mu$ are real and symmetric, at any point in
$\partial V$ the matrix $C^\mu n_\mu$ can be diagonalised with real
eigenvalues, and so the space ${\Bbb R}^N$ of dependent variables $\phi$ can
be written as the sum of the positive, negative and zero eigenspaces
of $C^\mu n_\mu$, or ${\Bbb R}^N=V_+\oplus V_-\oplus V_0$. Hence the
outward-pointing flux $j^\mu n_\mu=\phi^\dagger C^\mu n_\mu\phi$ can
be written as a term that is positive definite on $V_+$ plus one that
is negative definite on $V_-$. Integrating over $\partial V$, we can
then write (\ref{integratedbalancelaw}) schematically as
\begin{equation}
\int_V S\,dV=:||\text{out}||^2-||\text{in}||^2.
\end{equation}
In applications, we will split $||\text{out}||^2$ into
\begin{equation}
||\text{out}||^2=||\text{out}'||^2+||\text{out}''||^2,
\end{equation}
where $||\text{out}''||^2$ denotes any part of the outgoing flux that
we do not want to include in our estimate. We trivially obtain
\begin{equation}
\label{outinSdV}
||\text{out}'||^2\le ||\text{in}||^2+\int_V S\,dV.
\end{equation}

Now assume there is a slicing of $V$ by hypersurfaces of constant $t$
such that the system is symmetric hyperbolic with respect to $t$. Let
$c>0$ be the smallest eigenvalue of $C^t:=C^\mu(dt)_\mu$ anywhere in
$V$, and let $d$ be the largest positive eigenvalue of ${\cal D}$
anywhere in $V$, or zero if ${\cal D}$ is negative definite everywhere
in $V$. We have $S\le d\phi^\dagger\phi$ and $j^t\ge
c\phi^\dagger\phi$, and so we can bound the source term $S$ of the
balance law as
\begin{equation}
\label{Sfromjt}
S\le {d\over c} j^t.
\end{equation}

We slice $V$ into surfaces of constant $t$,
\begin{equation}
\int_V S\,dV=\int_{t_0}^{t_1}{\cal S}_t \,dt, \qquad {\cal
  S}_t:=\int_{V\cap\Sigma_t} S \,d\Sigma_t,
\end{equation}
where $t_1:=\sup_V t$ and $t_0:=\inf_V t$, and use the bound
(\ref{Sfromjt}) to obtain
\begin{equation}
{\cal S}_t\le {d\over c}{\cal E}_t, \qquad {\cal
  E}_t:=\int_{V\cap\Sigma_t} j^t \,d\Sigma_t.
\end{equation}
We now evaluate (\ref{outinSdV}) on the control volume
$V_t:=V\cap\{t'<t\}$ (the part of $V$ to the past of $t'=t$) to obtain
\begin{eqnarray}
{\cal  E}_{t}&=&||\text{in}||^2_{(\partial V)_t}
-||\text{out}||^2_{(\partial V)_t}
+\int_{t_0}^t{\cal S}_{t'}\,dt' \nonumber \\ \\
&\le&||\text{in}||^2+{d\over c}\int_{t_0}^t {\cal E}_{t'}\,dt',
\label{Etestimate}
\end{eqnarray}
where we have defined $(\partial V)_t:=\partial V\cap\{t'<t\}\subseteq
\partial V$. In (\ref{Etestimate}) we have used
$||\text{in}||^2_{\partial V_t}\le ||\text{in}||^2$, which follows
from $\partial V_t\subseteq \partial V$.

We now differentiate to turn the integral inequality
(\ref{Etestimate}) into a
differential one,
\begin{equation}
\label{dcalEdet}
{d\over dt}{\cal E}_{t}\le{d\over c}{\cal E}_{t}, \qquad
{\cal E}_{t_0}\le ||\text{in}||^2,
\end{equation}
and solve this to obtain
\begin{equation}
{\cal E}_{t}\le e^{{d\over c}(t-t_0)} ||\text{in}||^2.
\end{equation}
We then have
\begin{equation}
\label{intSdV}
\int_V S\,dV\le \int_{t_0}^{t_1}{d\over c}{\cal E}_t \,dt
\le \left(e^{{d\over c}(t_1-t_0)}-1\right) ||\text{in}||^2 , 
\end{equation}
and substituting (\ref{intSdV}) into (\ref{outinSdV}) we obtain the
desired estimate
\begin{equation}
\label{generalestimate}
||\text{out}'||^2\le  e^{{d\over c}(t_1-t_0)}||\text{in}||^2,
\end{equation}
as derived in \cite{Frittelli2005b}, but here for arbitrary $t_1-t_0$ and $V$.


\subsection{The scalar wave equation on Schwarzschild}
\label{section:1}


In spherical polar coordinates, symmetry of the matrices $C^\mu$ is
less obvious than it is in Cartesian coordinates . This is best
illustrated if we consider the scalar wave equation on the
Schwarzschild background, which decouples to linear order from the
metric perturbations, as already noted. It is
\begin{equation}
-2\psi_{,ur}-{2\over r}\psi_{,u}+{\cal A}(\psi_{,rr}+{2\over
  r}\psi_{,r})+{2m\over r^2}\psi_{,r} +{1\over
  r^2} \nabla^a\nabla_a\psi=0.
\end{equation}
Here and in the following, we write all equations in covariant form
with respect to the coordinates $x^a$ on $S^2$. Following
\cite{Frittelli2004}, we denote by $q_{ab}$ the abstract round unit
metric on $S^2$, by $q^{ab}$ its inverse, and by $\nabla_a$ the
covariant derivative with respect to $q_{ab}$, so
$\nabla_aq_{bc}=0$. Note that $\nabla_a$ commutes with partial
$\partial_u$ and $\partial_x$.

Following \cite{Frittelli2004}, we define the reduction variables
\begin{equation}
P:=(r\psi)_{,r}, \qquad Q_a:=\nabla_a\psi,
\end{equation}
and obtain the first-order system
\begin{eqnarray}
\label{waveeqnP}
2P_{,u}-{\cal A}P_{,r}  -{1\over r}\nabla^aQ_a&=&{2m\over r^2}(P-\psi), \\
\label{waveeqnQ}
Q_{a,r}-{1\over r}\nabla_a P&=&-{1\over r}Q_a, \\
\label{waveeqnpsi}
\psi_{,r}&=&{1\over r}(P-\psi).
\end{eqnarray}

Let $x^a=(\theta,\varphi)$ be the usual coordinates on $S^2$, in terms
of which $q_{ab}=\text{diag}(1,\sin^2\theta)$. Then
\begin{equation}
\nabla^a\nabla_a\psi=\psi_{,\theta\theta}+\cot\theta\,\psi_{,\theta}
+{1\over\sin^2\theta}\psi_{,\varphi\varphi}.
\end{equation}
To make the non-diagonal matrices $C^\theta$ and $C^\varphi$
symmetric, we need to expand the covector $Q_a$ in components with
respect to the non-coordinate, orthonormal basis
\begin{equation}
\label{S2frame}
\partial_\theta, \quad \partial_{\hat\varphi}:={1\over \sin\theta}\,\partial_\varphi.
\end{equation}
The system becomes 
\begin{eqnarray}
2P_{,u}-{\cal A}P_{,r}&& \nonumber \\
-{1\over r}\left(Q_{\theta,\theta}
+{1\over\sin\theta}Q_{{\hat\varphi},\varphi}\right)&=&
{2m\over r^2}(P-\psi)+{1\over r}\cot\theta\,Q_\theta, \nonumber \\  
\label{waveeqnPthetahatvarphi} \\
\label{waveeqnQtheta}
Q_{\theta,r}-{1\over r}P_{,\theta}&=&-{1\over r}Q_\theta, \\
\label{waveeqnQhatvarphi}
Q_{{\hat\varphi},r}-{1\over r\sin\theta} P_{,\varphi}&=&-{1\over r}Q_{\hat\varphi}, \\
\label{waveeqnpsicopy}
\psi_{,r}&=&{1\over r}(P-\psi).
\end{eqnarray}
The matrices $C^u$ and $C^r$ are diagonal and $C^\theta$ and
$C^\varphi$ are now symmetric. Replacing the coordinate $\theta$ by
$y:=-\cos\theta$ gets rid of the $\cot\theta$ term in $D$, see
\cite{Frittelli2004} ($s$ there is $-y$ here).

However, a more elegant approach to both establishing
  the symmetry of the $C^a$ and keeping track of Christoffel terms
  from covariant derivatives is to keep the equations covariant on
$S^2$. From (\ref{waveeqnP}-\ref{waveeqnpsi}) we read off
\begin{eqnarray}
j^u&=&2P^2, \\ 
j^r&=&-{\cal A}P^2+Q^aQ_a+\psi^2, \\ 
j^a&=&-{2\over
  r}PQ^a, \\ 
S&=&{4m\over r^2}P^2-{2\over r}(Q_aQ^a+\psi^2)+{A\over
  r}P\psi. 
\end{eqnarray}
It is easy to check that
\begin{equation}
\label{conservedcurrentexpanded}
{j^u}_{,u}+{j^r}_{,r}+\nabla_aj^a=S
\end{equation}
holds if and only if (\ref{waveeqnP}-\ref{waveeqnpsi}) hold, where
$\nabla_aj^a$ is the covariant divergence. This means that our
integration measure $dV$ must contain the covariant measure $d\Omega$
on the round two-sphere, or $d\Omega=\sin\theta\,d\theta\,d\varphi$ in
the standard coordinates.


\subsection{The vacuum Einstein equations, linearised in Bondi gauge
  about Schwarzschild}
\label{section:2}


In \cite{Frittelli2004}, the metric is written as
\begin{eqnarray}
\label{Frittellimetric}
ds^2&=&-{e^{2\beta}V\over r}du^2-2e^{2\beta} du\,dr \nonumber \\
&&+r^2 h_{ab}(dx^a-U^a\,du)(dx^b-U^b\,du).
\end{eqnarray}
Keeping in mind that $y=-\cos\theta$ and $S:=1-y^2=\sin^2\theta$, we
read off the identifications of the metric components of
Secs.~\ref{section:hyperbolicity} and \ref{section:Frittelli} given in
Table~\ref{table:nonlin}. 

\begin{table}
\setlength{\tabcolsep}{9pt} 
\renewcommand{\arraystretch}{1.5} 
\begin{tabular}{c|c|c}
Sec.~\ref{section:Frittelli} & Sec.~\ref{section:hyperbolicity} &  Schwarzschild\\
\hline
$r$ & $R=x=r$ & $r$ \\
$\beta$ & ${1\over 2}\ln G$ & $0$ \\
$V$ & ${rH\over G}$ & $r-2m$  \\
\hline
$h_{\theta\theta}$ & $e^{2Sf}$ & $1$ \\
$h_{\varphi\varphi}$ & $Se^{-2Sf}$ & $S$ \\
$h_{\theta\varphi}$ & $0$ & $0$ \\
\hline
$U^\theta$ & $-\sqrt{S}b$ & $0$ \\ 
$U^\varphi$ & $0$ & $0$ \\
\end{tabular}
\caption{Comparison of our notation for the nonlinear variables in
  Secs.~\ref{section:hyperbolicity} and \ref{section:Frittelli}.}
\label{table:nonlin}
\end{table}

\begin{table}
\setlength{\tabcolsep}{9pt} 
\renewcommand{\arraystretch}{1.5} 
\begin{tabular}{c|c|c}
Sec.~\ref{section:Frittelli} & Sec.~\ref{section:hyperbolicity} & toy model \\
\hline
$\tilde \beta$ & ${1\over 2}\delta G$ & ${1\over 2}\delta G$ \\
$\tilde v$ & $\delta H-\delta G$ 
& ${r^2\over 2}\overline{\delta H}-r\delta G$\\
\label{tildeheqn}
$\tilde h_{\theta\theta}$ & $2S\delta f$ & $\overline{\delta f}$ \\
$\tilde u_\theta$ & $-r\sqrt{S}\delta b$ & $-\overline{\delta b}$ \\
$P_{\theta\theta}$ & $2S(r\delta f)_{,r}$ & $r\overline{\delta
  f}_{,x}$ \\
$Q_\theta$ & $-\sqrt{S}(r^2\delta b)_{,r}-\sqrt{S}\delta G_{,y}$ & 
$-rQ$ \\
$T_\theta$ & ${1\over 2}\sqrt{S}\delta G_{,y}$ &
 ${r\over 2}T$ \\
$J_\theta$ & $\sqrt{S}(2S\delta f)_{,y}-\sqrt{S}(r^2\delta b)_{,r}$ &
$rJ$ \\
\end{tabular}
\caption{Comparison of our notation for the linear perturbations in
  Secs.~\ref{section:hyperbolicity} and \ref{section:Frittelli}. We
  only list the independent perturbations present in twist-free
  axisymmetry, see
  (\ref{tildehvarphivarphiconstraint},\ref{Pvarphivarphiconstraint})
  for $\tilde h_{\varphi\varphi}$ and $P_{\varphi\varphi}$.}
\label{table:pert}
\end{table}

We denote the perturbations of $V$, $\beta$, $h_{ab}$ and $U^a$
about the vacuum Schwarzschild solution by
$\tilde V$, $\tilde \beta$, $\tilde h_{ab}$ and $\tilde U^a$.  (In
contrast to \cite{Frittelli2005a}, we have added the tildes on the
perturbations $\tilde\beta$ and $\tilde U^a$ to distinguish them
from the full variables.) We also replace
  $\tilde U_a$ and $\tilde V$ by 
\begin{eqnarray}
\tilde u_a&:=&r\tilde U_a, \\
\tilde v&:=&{\tilde V\over r}.
\end{eqnarray}
This makes all variables dimensionless, and therefore all lower-order terms
become proportional to $2m/r^2$ or $1/r$. Our perturbation variables
are summarised in Table~\ref{table:pert}.

We introduce the reduction
variables of \cite{Frittelli2005a}, which are
\begin{eqnarray}
\label{Pabdef}
P_{ab}&:=&(r\tilde h_{ab})_{,r}, \\
\label{Qadef}
Q_a&:=&(r\tilde u_a)_{,r}-2\tilde\beta_{,a}, \\
\label{Jadef}
J_a&:=&\nabla^b\tilde h_{ab}+(r\tilde u_a)_{,r}, \\
\label{Tadef}
T_a&:=&\tilde \beta_{,a}.
\end{eqnarray}
These comprise 8 first derivatives of the 6 metric perturbations.

Note that the linearisation of the Bondi gauge condition $\det
h_{ab}=\det q_{ab}$ is $q^{ab}\tilde h_{ab}=0$, or in coordinates,
\begin{equation}
\label{tildehvarphivarphiconstraint}
\tilde h_{\varphi\varphi}=-\sin^2\theta\, \tilde h_{\theta\theta}.
\end{equation}
By definition, $P_{ab}$ is also trace-free, and so
\begin{equation}
\label{Pvarphivarphiconstraint}
P_{\varphi\varphi}=-\sin^2\theta\, P_{\theta\theta}.
\end{equation}

To complete the system with an evolution equation for $\tilde v$ while
maintaining the symmetric hyperbolic form, we introduce the further
variables
\begin{eqnarray}
\label{calPadef}
{\cal P}_a&:=&\nabla^bP_{ab}, \\
{\cal Q}&:=&\nabla^aQ_a, \\
\hat{\cal Q}&:=&\epsilon^{ab}\nabla_aQ_b, \\
{\cal J}&:=&\nabla^aJ_a, \\
 \hat{\cal J}&:=&\epsilon^{ab}\nabla_aJ_b, \\
{\cal U}&:=&\nabla^a\tilde u_a, \\
\hat{\cal U}&:=&\epsilon^{ab}\nabla_a\tilde u_b, \\
{\cal T}&:=&\nabla^aT_a. 
\label{calTdef}
\end{eqnarray}
These comprise 2 additional first derivatives of the metric (namely
$\cal U$ and $\hat{\cal U}$, for a total of 10) and 7 second
derivatives of the metric.  We move tensor
indices $a,b,...$ on $S^2$ with $q_{ab}$ and $q^{ab}$, and note that
this commutes with taking derivatives in $u$ and $r$. $\epsilon_{ab}$
is the volume form on the unit 2-sphere, with defining properties
\begin{equation}
\nabla_a\epsilon_{bc}=0, \qquad \epsilon_{ac}{\epsilon^b}_c=q_{ab}.
\end{equation}

We can decompose the vector field $Q_a$ in terms of potentials $Q$ and
$\hat Q$ that are determined (up to a constant) as solutions of the
Poisson equation on the unit 2-sphere, as follows:
\begin{eqnarray}
\label{Qadecomposition}
{\cal Q}_a&=&\nabla_aQ-{\epsilon_a}^b\nabla_b\hat Q, \\
\nabla^a\nabla_aQ&=&{\cal Q}, \\
\nabla^a\nabla_a\hat Q&=&\hat{\cal Q}.
\end{eqnarray}
We can use (\ref{Qadecomposition}) to show that
\begin{eqnarray}
&&\nabla^b\left(\nabla_aQ_b+\nabla_bQ_a-q_{ab}\nabla^cQ_c\right)
  \nonumber \\
&=& \nabla^b\nabla_b{\cal Q}_a+Q_a \nonumber \\
&=&\nabla_a{\cal Q}-{\epsilon_a}^b\hat{\cal Q}+2Q_a,
\label{QaLaplace}
\end{eqnarray}
where in both lines we have used
\begin{equation}
(\nabla_a\nabla_b-\nabla_b\nabla_a)Q^b=-R_{ab}Q^b=-Q_a,
\end{equation}
with $R_{ab}=q_{ab}$ the Ricci tensor on the unit round 2-sphere. 
The potentials $Q$ and $\hat Q$ were introduced only to
derive (\ref{QaLaplace}), and are not part of our system.

The evolution equations for the reduction variables already introduced
in \cite{Frittelli2005a} are
\begin{eqnarray}
\label{Pabeqncorrected}
2P_{ab,u}-{\cal A} P_{ab,r} && \nonumber \\
+{1\over r}\left(2\nabla_{(a}Q_{b)}-q_{ab}\nabla^cQ_c\right) &=&
{2m\over r^2}(P_{ab}-\tilde h_{ab}), \\
Q_{a,r}+{1\over r}\nabla^bP_{ab}&=&{1\over r}(J_a-Q_a-2T_a+2\tilde u_a), 
\nonumber \\ 
\label{Qaeqncorrected} \\
\label{Jaeqncorrected}
J_{a,r}&=&{1\over r}(2\tilde u_a-4T_a), \\
\tilde u_{a,r}&=&{1\over r}(Q_a+2T_a-u_a). \nonumber \\ 
\label{Uaeqn}\\
\label{Taeqn}
T_{a,r}&=&0, \\
\tilde h_{ab,r}&=&{1\over r}(P_{ab}-\tilde h_{ab}), \\
\tilde\beta_{,r}&=&0.
\end{eqnarray}
Eqs.~(\ref{Pabeqncorrected}-\ref{Jaeqncorrected}) incorporate minor
corrections of their counterparts in \cite{Frittelli2005a}, see
Appendix~\ref{appendix:Frittellicorrections}.  The evolution equations
for our additional variables are
\begin{eqnarray}
2{\cal P}_{a,u}-{\cal A}{\cal P}_{a,r} && \nonumber \\ +{1\over r}
(\nabla_a{\cal Q}-{\epsilon_a}^b\nabla_b\hat{\cal Q})&=&-{2\over r}Q_a
\nonumber \\
&&\hskip-1cm  +{2m\over r^2}({\cal P}_a-J_a+Q_a+2T_a),\label{calPaeqn}
\label{calPaeqn} \\
\label{calQeqn}
{\cal Q}_{,r}+{1\over r}\nabla^a{\cal P}_a&=&{1\over r}({\cal
  J}-{\cal Q}-2{\cal T}+2{\cal U}), \\
\label{calQhateqn}
\hat{\cal Q}_{,r}+{1\over r}\epsilon^{ab}\nabla_a{\cal
  P}_b&=&{1\over r}(\hat{\cal
  J}-\hat{\cal Q}+2\hat{\cal U}), \\
{\cal J}_{,r}&=&{1\over r}(2{\cal U}-4{\cal T}), \\
\hat{\cal J}_{,r}&=&{2\over r}\hat{\cal U}, \\
{\cal U}_{,r}&=&{1\over r}({\cal Q}+2{\cal T}-{\cal U}), \\
\hat{\cal U}_{,r}&=&{1\over r}(\hat{\cal Q}-\hat{\cal U}), \\
{\cal T}_{,r}&=&0, \label{calTeqn}\\
\tilde v_{,r}&=&{1\over r}\left({1\over 2}{\cal J}-{\cal T}+{\cal
  U}+2\tilde\beta-\tilde v\right). \nonumber \\
\label{Vtildeeqn}
\end{eqnarray}

The explicit matrices $C^\mu$ are given in
Appendix~\ref{appendix:Frittellicorrections}. For the more elegant
covariant-on-$S^2$ approach to symmetric hyperbolicity, we define
\begin{equation}
{X_{ab}}^{cd}:={1\over
  2}({q_a}^c{q_b}^d+{q_b}^c{q_a}^d-q_{ab}q^{cd}),
\end{equation}
the projection operator into the space of symmetric tracefree
2-tensors on $S^2$. We can then write
(\ref{Pabeqncorrected},\ref{Qaeqncorrected}) as
\begin{eqnarray}
2P_{ab,u}-{\cal A}P_{ab,r}+{2\over
  r}{X_{ab}}^{cd}\nabla_dQ_c&=&\text{l.o.}, \\\
Q_{c,r}+{1\over r}{X_{abc}}^d\nabla_d P_{ab}&=&\text{l.o.},
\end{eqnarray}
and the conserved current as
\begin{eqnarray}
\label{judef}
j^u&:=&2{\bf P}^\dagger{\bf P} \\
\label{jrdef}
j^r&:=&-{\cal A}{\bf P}^\dagger{\bf P}+{\bf Q}^\dagger{\bf Q} \\
j^d&:=&{2\over r}X^{abcd}P_{ab}Q_c+{1\over r}\left(q^{da}{\cal
  Q}+\epsilon^{da}\hat{\cal Q}\right)P_a, \\
S&=&{1\over r}(...)+{2m\over r^2}(...),
\end{eqnarray}
where we have defined the shorthands
\begin{eqnarray}
\label{PdaggerPdef}
{\bf P}^\dagger{\bf P}&:=&{1\over 2}P_{ab}P^{ab}+P_aP^a \\
&=&P_{\theta\theta}^2+P_{\theta{\hat\varphi}}^2+P_\theta^2+P_{\hat\varphi}^2, \\
\label{QdaggerQdef}
{\bf Q}^\dagger{\bf Q}&:=&Q_aQ^a+...=Q_\theta^2+Q_{\hat\varphi}^2+...,
\end{eqnarray}

The factor of $1/2$ in front of $P_{ab}P^{ab}$ compensates for
double-counting of its algebraically independent components, see
(\ref{Pvarphivarphiconstraint}). The balance
law(\ref{conservedcurrentexpanded}) holds if and only if our
first-order reduction of the linearised Einstein equations holds.

For brevity, we have not written out $S$ in full.
Our introduction of $\tilde u_a$ and $\tilde v$ has the advantage,
relative to \cite{Frittelli2005a}, that in the Minkowski case $m=0$
${\cal D}$ becomes $1/r$ times a matrix of integers, so all its
eigenvalues take the form $\lambda_i(r)=\bar\lambda_i/r$, and $d=\bar
d/r_0$, where $\bar d$ is the largest $\lambda_i$. 

Looking at the evolution equations, in order to add $\tilde v$ to our
system and then close it as first-order symmetric hyperbolic system we
have effectively introduced a system of variables and equations
that duplicates Frittelli's original system, but at one derivative
higher. Geometrically, this is the level of curvature, rather than of
the connection. However, our estimate includes, beside the 6 metric
perturbations, only 10 first and 7 second derivatives of the metric,
far short of the full set of Ricci rotation coefficients and null
curvature components. These are listed in
Table~\ref{table:estimatevariables}.

At both the level of the connection and the level of curvature, we
have in effect two pairs of wave equations, but in
different objects, $(P_{ab},Q_a)$, with ${P_a}^a=0$, and ${\cal
  P}_a,{\cal Q},\hat{\cal Q}$. Using (\ref{QaLaplace}) in
(\ref{calPaeqn}) was essential for bringing the pair
(\ref{calPaeqn},\ref{calQeqn}) into a form similar to
(\ref{Pabeqncorrected},\ref{Qaeqncorrected}), with the same matrices
$C^\mu$.

As the background solution is spherically symmetric, all perturbations
can be split into polar and axial parts, where the axial perturbations
change sign under a reflection of $S^2$, or reflection in space. All
genuine scalars on $S^2$, in our case $\psi$, $\tilde V$ and $\tilde
\beta$ are polar. Vectors on $S^2$ can be split into polar and axial
parts as in (\ref{Qadecomposition}) for the example of
$Q_a$. Tracefree symmetric tensors can be split as, for example,
\begin{equation}
P_{ab}=\left(2\nabla_{(a}\nabla_{b)}-q_{ab}\nabla^c\nabla_c\right)P
+2{\epsilon_{(a}}^c\nabla_{b)}\nabla_c \hat P,
\end{equation}
in terms of a scalar $P$ and pseudoscalar $\hat P$.  Axial parts are
hatted. Then axial and polar perturbations decouple from each other.

In twist-free axisymmetry, only the polar
perturbations are present. We note in passing that substituting the solution
$\tilde\beta=T_a=0$ of $\tilde\beta_{,r}=0$, and restricting the
background solution to the Minkowski spacetime by setting $m=0$, makes
Eqs.~(\ref{Pabeqncorrected},\ref{Qaeqncorrected}) equivalent to
Eqs.~(D25,D26) of Paper I.

We could also write the linear perturbation equations in terms of
scalars and pseudoscalars only, in order to explicitly decouple polar
and axial perturbations. This form of the system would not be
symmetric hyperbolic, as all angular derivatives would appear in the
form $\nabla^c\nabla_c$, but it could be made symmetric hyperbolic
again by re-introducing first angular derivatives as reduction
variables, as for the scalar wave equation. However, this would mean
duplicating all vectors, for example defining $Q_a=\nabla_aQ$ to be a
true vector (polar) and adding $\hat Q_a=\epsilon_{ab}\nabla^b\hat Q$
as a pseudovector (axial).

Looking back, to merely write the second-order Einstein equations
first-order form with a minimum number of variables, one already has
to introduce all the reduction variables (first derivatives of the
metric) (\ref{Pabdef}-\ref{Tadef}). This is true if $\tilde v$ is
included or not. There is one reduction constraint
\begin{equation}
\nabla^bP_{ab}=(r\nabla^b\tilde h_{ab})_{,r}=\left(r(J_a-Q_a-2T_a)\right)_{,r}.
\end{equation}
The second derivative $(r\nabla^b\tilde h_{ab})_{,r}$ that can be thus
written in two ways appears in only one place, and so there is a
one-parameter family of first-order reductions with inequivalent
principal parts. It turns out this contains the symmetric hyperbolic
first-order form of the equations with the equation for $\tilde v$
excluded found by Frittelli. To bring the full system, with $\tilde v$
included, into a first-order symmetric hyperbolic form, we had to
further add all the variables (\ref{calPadef}-\ref{calTdef}).

\begin{table}
\begin{tabular}{c|c|c}
& ${\bf P}$ & ${\bf Q}$ \\
\hline
$g$ & --- &  
$\tilde\beta,\, 
\tilde h_{\theta\theta},\, 
\tilde h_{\theta{\hat\varphi}},\,
\tilde u_a;\,\tilde v$ \\
$\partial_r g$ & $(r\tilde h_{ab})_{,r}$ & 
$(r\tilde u_a)_{,r}$ \\
$\nabla g$ & --- &
$\tilde\beta_{,a},\, 
\nabla^b\tilde h_{ab};\,  
\nabla^a \tilde u_a,\, 
\epsilon^{ab}\nabla_a\tilde u_b$ \\
$\partial_r\nabla g$ & $\nabla^a(r\tilde h_{ab})_{,r},$ & 
$\nabla^a(r \tilde u_a)_{,r},\, 
\epsilon^{ab}\nabla_a (r\tilde u_b)_{,r}$ \\
$\nabla\nabla g$ & --- & 
$\nabla^a\nabla_a\tilde \beta,\,
\nabla^a\nabla^b\tilde h_{ab},\,
{\epsilon_c}^b\nabla^a\nabla^c\tilde h_{ab}$ \\
\end{tabular}
\caption{List of the quantities involved in our $L^2$ estimates,
  written out in terms of metric perturbations and their derivatives,
  and ordered by derivative of the metric and by left-moving variables
  ${\bf P}$ and right-moving variables ${\bf Q}$. The quantities in
  the first line before the semicolon, the second line, and the third
  line before the semicolon were already introduced as variables in
  \cite{Frittelli2005a}. We have introduced the remaining quantities
  as variables to obtain a symmetric hyperbolic first-order system
  including all metric perturbations.}
\label{table:estimatevariables}
\end{table}
 

\subsection{Estimates for characteristic initial value and
  initial-boundary value problems}
\label{section:specificestimates}


We now use the symmetric hyperbolic form of the linearised Einstein
equations in Bondi gauge to obtain energy estimates on control volumes
of interest. We introduce the Schwarzschild time coordinate
\begin{equation}
t:=u+r_*\quad\Rightarrow\quad dt=du+dr_*=du+{\cal A}^{-1}dr,
\end{equation}
where $r_*$ is the usual tortoise radius, and where we have defined the
shorthand
\begin{equation}
{\cal A} :=1-{2m\over r}.
\end{equation}
Following \cite{Frittelli2005a}, we observe that
\begin{equation}
C^t:=C^\mu(dt)_{,\mu}=C^u+{\cal A}^{-1}C^r,
\end{equation}
where $C^u$ and $C^r$ are explicitly given in
(\ref{Cudef},\ref{Crdef}), is positive definite on ${\Bbb R}^N$ with
smallest eigenvalue $1$ for our system, independently of $r$, so for
the smallest eigenvalue anywhere in $V$ we have $c=1$. Equivalently,
with $j^u$ and $j^r$ given by (\ref{judef}) and (\ref{jrdef}), we have
\begin{equation}
j^t:=j^u+{\cal A}^{-1}j^r=
{\bf P}^\dagger{\bf P}+{\cal A}^{-1}{\bf Q}^\dagger{\bf Q}, \label{jtdef}
\end{equation}
where ${\bf P}^\dagger{\bf P}$ and ${\bf Q}^\dagger{\bf Q}$ were
defined in (\ref{PdaggerPdef}-\ref{QdaggerQdef}). 

\begin{figure}
\includegraphics[scale=0.3]{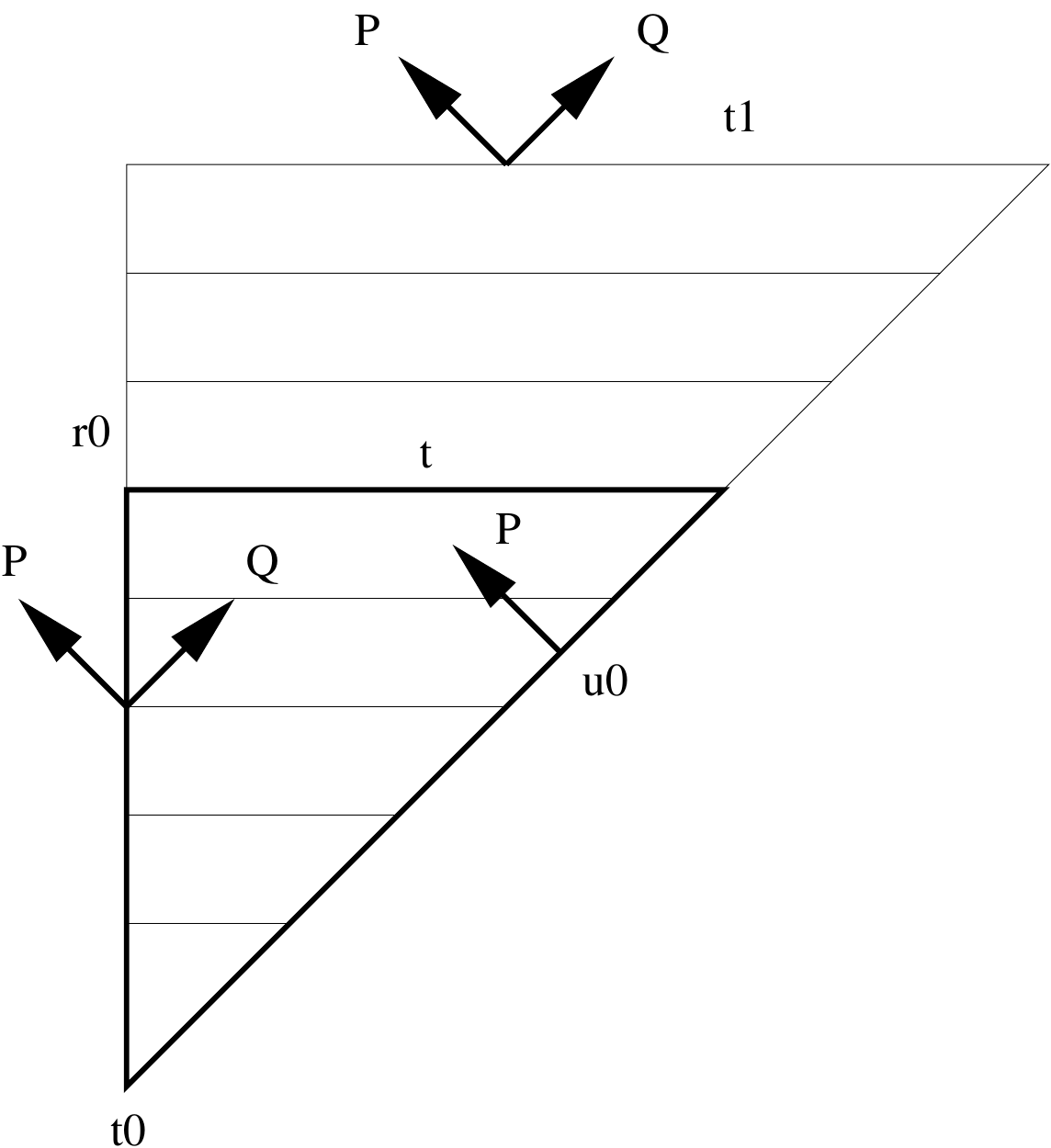}
\caption{Spacetime diagram of the control volume $V_1$, reduced by
  spherical symmetry, so that every point in the plot corresponds to a
  spacelike 2-sphere. Ingoing and outgoing spherical null surfaces are
  lines at 45 degrees. Horizontal lines represent surfaces of constant
  $t$, by which we slice $V_1$. $V_1$ is bounded by $r=r_0$ (left),
  $u=u_0$ (bottom right) and $t=t_1$ (top). The thicker line shows the
  volume $V_t$ bounded by $t=t'<t_1$, which is needed for the estimation of
  ${\cal E}_t$ in (\ref{Etestimate}).  Arrows labelled ${\bf P}$ and
  ${\bf Q}$ symbolise energy fluxes.The estimate on $V_1$ is
  (\ref{V1estimate}).}
\label{fig:V1}
\end{figure}

The estimates in \cite{Frittelli2005a,Frittelli2004,Frittelli2005b} are
on the control volume $V_1$ shown in Fig.~\ref{fig:V1}: the product of $S^2$
with the triangle bounded by $u=u_0$, $r=r_0$, and $t=t_1$. We have
\begin{equation}
-\int\displaylimits_{u=u_0}j^u-\int\displaylimits_{r=r_0}j^r
+\int\displaylimits_{t=t_1}j^t=\int_{V_1} S.
\end{equation}
The signs come from $du$ at $u=u_0$ and $dr$ at $r=r_0$ pointing into
$V$, and $dt$ at $t=t_1$ pointing out. We will ignore the flux out of
$r=r_0$ (as is standard practice for initial-boundary value problems),
and so
\begin{eqnarray}
||\text{in}||^2&=&\int\displaylimits_{u=u_0}2{\bf P}^\dagger{\bf P}
+\int\displaylimits_{r=r_0}{\bf Q}^\dagger{\bf Q},\\
||\text{out}'||^2&=&\int\displaylimits_{t=t_1}
{\bf P}^\dagger{\bf P}+{\cal A}^{-1}{\bf Q}^\dagger{\bf Q} \\
||\text{out}''||^2&=&\int\displaylimits_{r=r_0}{\cal A}{\bf P}^\dagger{\bf P}.
\end{eqnarray}
Hence, from the general formula (\ref{generalestimate}), and with
$c=1$, our estimate is
\begin{eqnarray}
&&\int\displaylimits_{t=t_1} \left(
{\bf P}^\dagger{\bf P}+{\cal A}^{-1}{\bf Q}^\dagger{\bf Q}\right) \nonumber \\
&\le& e^{d(r_0)\,(t_1-t_0)} \left(
\,\int\displaylimits_{u=u_0}2{\bf P}^\dagger{\bf P}
+\int\displaylimits_{r=r_0}{\bf Q}^\dagger{\bf Q} \right).
\label{V1estimate}
\end{eqnarray}
Recall that $d$ is defined as the largest positive eigenvalue of
${\cal D}$ in $V$. All elements of ${\cal D}$ are constants times
$1/r$ or $2m/r^2$, so $d$ depends only on $r_0$, the smallest
value of $r$ in $V$, and we have written $d=d(r_0)$ to emphasise this.
On the Minkowski background, the $2m/r^2$ terms are absent, and so
$d=\bar d/r_0$, where $\bar d$ is the largest eigenvalue of $r{\cal
  D}$ (with $m=0$). 

\begin{figure}
\includegraphics[scale=0.3]{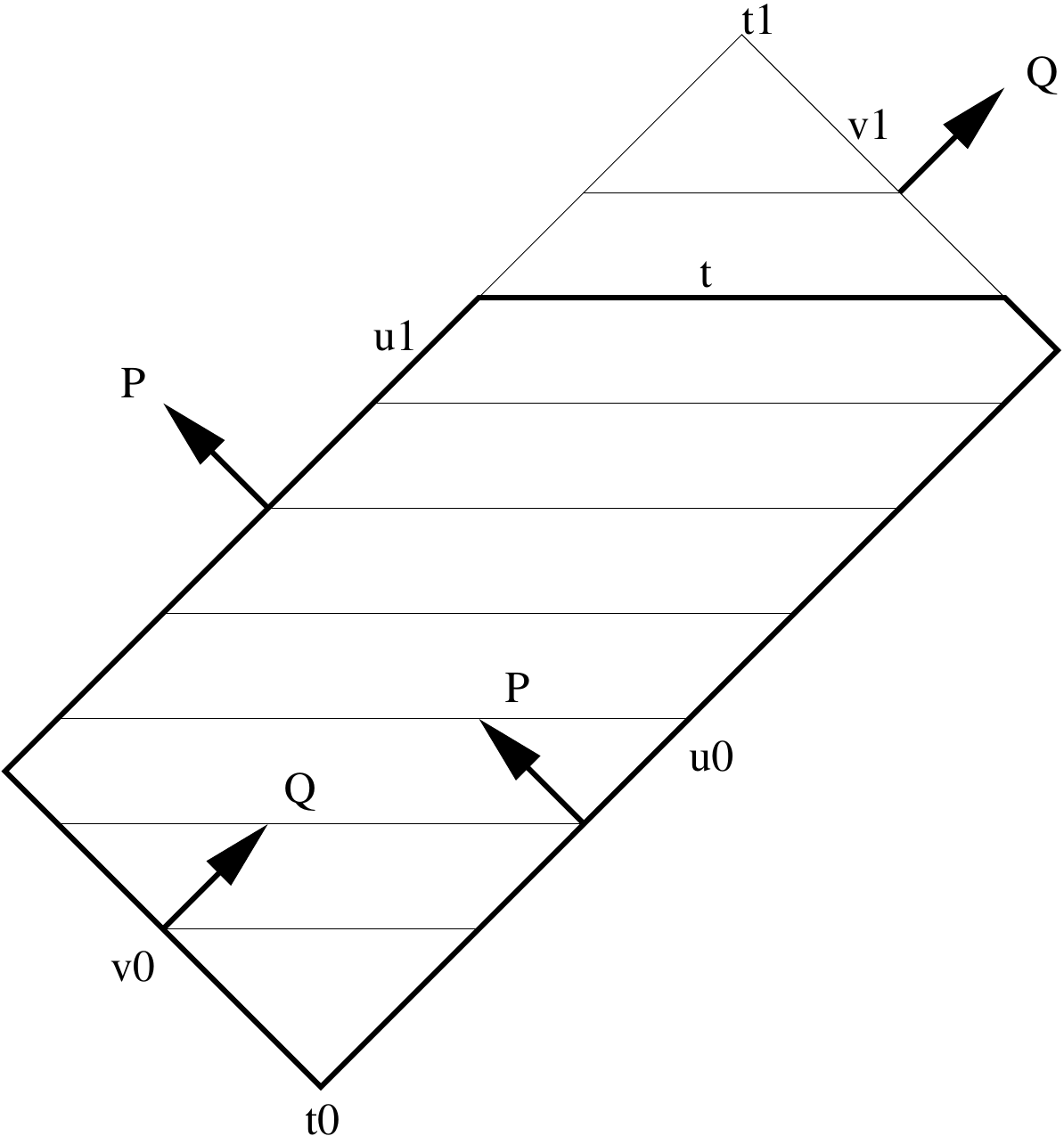}
\caption{ Spacetime diagram of the control volume $V_2$, bounded by
  $v=v_0$ (bottom left), $u=u_0$ (bottom right), $u=u_1$ (top left)
  and $v=v_1$ (top right). Otherwise as in Fig.~\ref{fig:V1}. The
  estimate on $V_2$ is (\ref{V2estimate}).}
\label{fig:V2}
\end{figure}

A second control volume of interest is shown in Fig.~\ref{fig:V2}: the product
of $S^2$ with the null rectangle triangle bounded by $u=u_0$ and
$v=v_0$, and $u=u_1$ and $v=v_1$. The null coordinate $v$ on the
Schwarzschild background is defined by
\begin{equation}
v:=u+2r_*\quad\Rightarrow\quad dv=du+2{\cal A}^{-1}dr,
\end{equation}
and hence 
\begin{equation}
j^v:=j^u+2{\cal A}^{-1}j^r=2{\cal A}^{-1}{\bf Q}^\dagger{\bf Q}. 
\end{equation}
The corresponding estimate is
\begin{eqnarray}
&&\,\int\displaylimits_{u=u_1}2{\bf P}^\dagger{\bf P}
+\int\displaylimits_{v=v_1}2{\cal A}^{-1}{\bf Q}^\dagger{\bf Q}
\nonumber \\
&\le& e^{d(r_0)\,(t_1-t_0)} \left(
\,\int\displaylimits_{u=u_0}2{\bf P}^\dagger{\bf P}
+\int\displaylimits_{v=v_0}2{\cal A}^{-1}{\bf Q}^\dagger{\bf Q}
\right).
\label{V2estimate}
\end{eqnarray}

\begin{figure}
\includegraphics[scale=0.3]{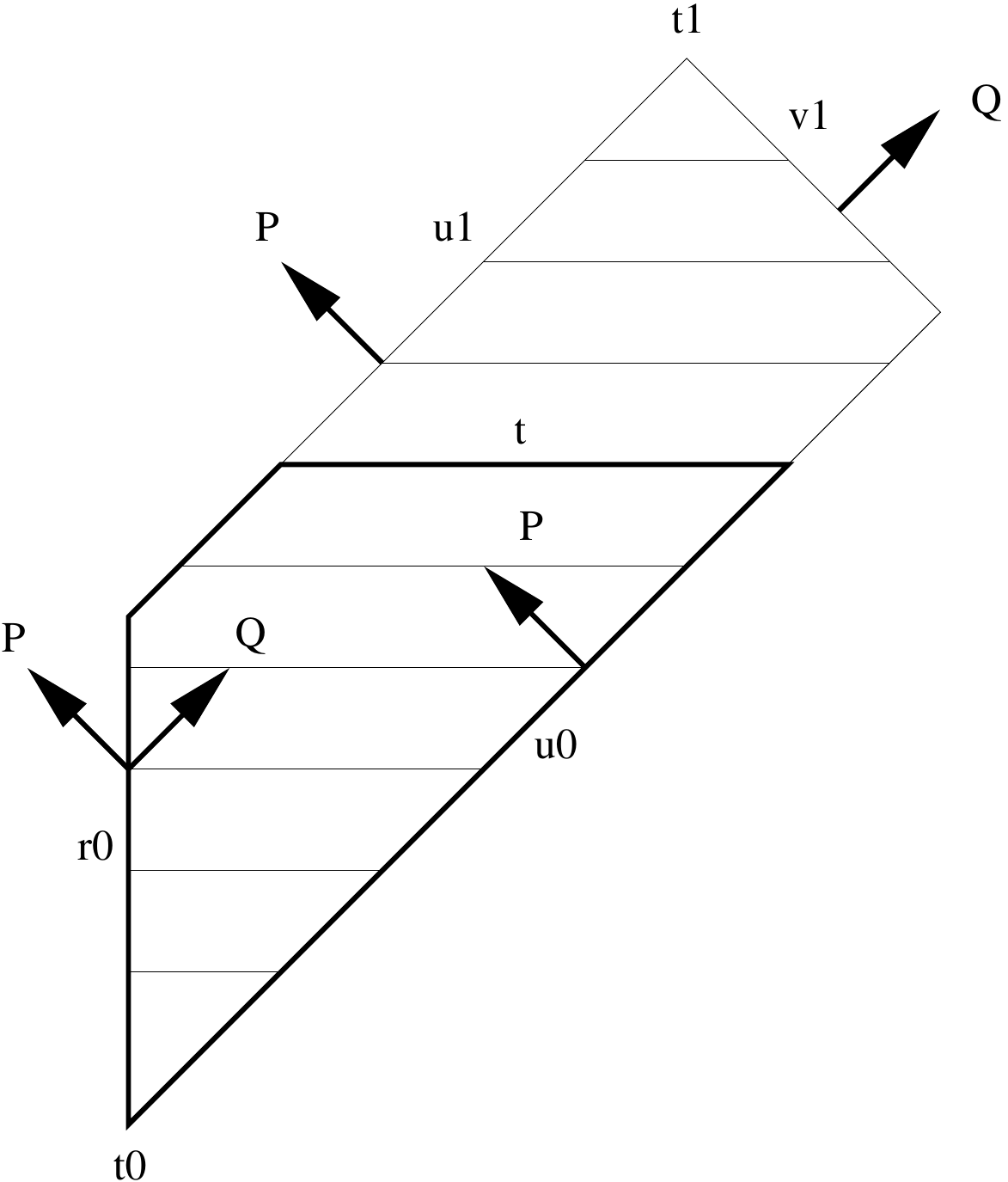}
\caption{ Spacetime diagram of the control volume $V_3$, bounded by
  $r=r_0$ (left), $u=u_0$ (bottom right), $u=u_1$ (top left) and
  $v=v_1$ (top right). Otherwise as in Fig.~\ref{fig:V1}. The estimate
  on $V_3$ is (\ref{V3estimate}).}
\label{fig:V3}
\end{figure}

A third control volume of interest is shown in Fig.~\ref{fig:V3}: the
product of $S^2$ with the null right trapezoid bounded by $u=u_0$, $r=r_0$, and
$u=u_1$ and $v=v_1$. Again we will ignore the flux out of $r=r_0$. The
corresponding estimate is
\begin{eqnarray}
&&\,\int\displaylimits_{u=u_1}2{\bf P}^\dagger{\bf P}
+\int\displaylimits_{v=v_1}2{\cal A}^{-1}{\bf Q}^\dagger{\bf Q}
\nonumber \\
&\le& e^{d(r_0)\,(t_1-t_0)} \left(
\,\int\displaylimits_{u=u_0}2{\bf P}^\dagger{\bf P}
+\int\displaylimits_{r=r_0}{\bf Q}^\dagger{\bf Q} \right).
\label{V3estimate}
\end{eqnarray}

A hypothetical fourth estimate would be (\ref{V3estimate}) with $m=0$
and $r_0=0$, that is the pure null initial-value problem on an
outgoing cone with regular vertex at $r=0$. (We need to set $m=0$ so the
background has a regular centre). This is the PDE problem we have
been considering in Papers~I and II in this series. Regularity on the
central worldline $r=0$ requires that all metric variables take their
Minkowski values. Hence their perturbations, and in particular the
right-moving perturbations ${\bf Q}$ vanish at $r=0$. Independently,
$r=0$ is no longer a boundary. Hence the last term in
(\ref{V3estimate}) would be absent.

However, we recall that on Minkowski spacetime $d(r_0)=\bar d/r_0$,
with $\bar d\simeq 5.0>0$ . Hence $\exp d(r_0)\,(t_1-t_0)$ grows
arbitrarily rapidly with $t$ as $r_0\to 0$, so the limit $r_0\to 0$ of
(\ref{V3estimate}) as written does not exist. This problem already
arises for the scalar wave equation on Minkowski. In
Appendix~\ref{appendix:tricks} we attempt some simple ways around this
problem, and show that one of them works for the scalar wave equation
on Minkowski, but none work for the metric perturbations. 
In Appendix~\ref{appendix:numericalexperiments} we carry out some
numerical tests of the hypothetical estimate on a regular null cones.


\section{Conclusions}
\label{section:conclusions}


In the Introduction, we motivated the desirability of a well-posedness
proof for formulations of the Einstein equations on null cones, not
just geometrically, but for specific formulations of the Einstein
equations that are used in numerical relativity. These formulations
use ``Bondi-like'' coordinates, where coordinate lines of constant
$(u,\theta,\varphi)$ are outgoing null rays, and where we evolve only
free data and solve ODEs along the null rays to reconstruct the full
metric on each time slice.

One of these formulations, using Bondi coordinates, has been used
successfully for Cauchy-characteristic matching \cite{WinicourLRR2012}
in full generality, see for example \cite{Maetal2023}. This and the
incomplete well-posedness result of \cite{Frittelli2005a} appeared to
be in tension with recent results \cite{GHZ2020,GBHPZ2022,GBHPZ2022}
that showed that minimal first-order reductions of this and similar
formulations are not strongly hyperbolic, and which we have verified
in Sec.~\ref{section:hyperbolicity}.

Our tentative resolution is that the characteristic initial-boundary
value problem is well-posed in $L^2$ of the metric plus {\em selected}
first and second derivatives. This is the ``skewed'' norm whose
existence was conjectured in \cite{GHZ2020,GBHPZ2022,GBHPZ2023}. In
Sec.~\ref{section:Frittelli}, we have proved this for the
linearisation of Bondi gauge about Schwarzschild, for the null
initial-boundary value problem and the double null initial value
problem. 

Based on this proof, we conjecture that 1) the initial-boundary value
problem is well-posed for the linearisation about an arbitrary
background; 2) this holds for any Bondi-like gauge; 3) this holds also
for the initial value-problem on a null cone with regular vertex.

We expect that generalising the symmetric hyperbolic form of the
linearised Einstein equations to an arbitrary background solution can
be done because we expect all additional terms to be of lower order.
The existence of a symmetric hyperbolic form of the linearised
Einstein equations in other gauges seems plausible because they can be
written as two wave equations coupled to transport equations along the
outgoing null cones, coupled only through lower-order terms.

The third conjecture appears to be the most challenging.
Unfortunately, the methods of \cite{Frittelli2004,Frittelli2005a} that
we have used for the estimates above do not allow an estimate for the
initial value problem with initial data on a regular null cone.
Chrusciel has given an existence proof for solutions of the Einstein
equations with initial data specified on a null cone with regular
vertex \cite{Chrusciel2013}. This uses harmonic coordinates, relying
on results of Dossa \cite{Dossa1997} for quasilinear wave
equations. Those proofs suggest that we may need to split off the
lowest powers of $r$ and corresponding lowest spherical harmonics
$Y_{lm}$ as an approximate solution (resulting in a polynomial in $x$
and $y$), and control only the remainder in an energy norm.

For a proof of well-posedness of Cauchy-characteristic {\em matching}
along the lines of the present paper to succeed, the set of variables
that are passed between the Cauchy code and the null code probably has
to coincide with the set of variables that appear at the matching
surface in the separate well-posedness estimates on both sides.

Setting aside the difficulties with proving well-posedness, one may
wonder if the linearised Einstein equations in Bondi coordinates on
null cones emanating from a regular centre are actually already
well-posed in the norm $\int{\bf P}^\dagger{\bf P}\,dS$ on those
null cones, and only our estimates for the lower-order terms are not
sharp enough to see this. Numerical experiments with the code of
\cite{axinull_formulation,axinull_critscalar} are described in
Appendix~\ref{appendix:numericalexperiments}. While these cannot of
course prove stability, they seem to be compatible with it.


\acknowledgments

The author is grateful to Jonathan Luk, David Hilditch, Thanasis
Giannakopoulos and Piotr Chrusciel for helpful discussions, and
particularly to TG for detailed comments on the manuscript.


\appendix



\section{A textbook example of a weakly hyperbolic system}
\label{appendix:toysystem}


A textbook example of a PDE system that is weakly but not
strongly hyperbolic is given in \cite{KreissLorenz2004},~pp.~29-30: 
\begin{equation}
\phi_{,t}=\left(\begin{array}{cc}
\lambda & 1 \\ 0 & \lambda \\
\end{array}\right)\phi_{,x}, \qquad
\phi:=\left(\begin{array}{cc}
\phi_1 \\ \phi_2  \\
\end{array}\right).
\end{equation}
It has the general solution
\begin{equation}
\label{toymodelsolution}
\phi(t,x)=\int_{-\infty}^\infty \left[
a_{(k)}\left(\begin{array}{c} 1 \\ 0 \\
\end{array}\right)
+b_{(k)} \left(\begin{array}{c} ikt \\ 1 \\
\end{array}\right)\right]e^{ik(x+\lambda t)}\,dk.
\end{equation}
Clearly this is not well-posed in the $L^2$ norm
$\int(\phi_1^2+\phi_2^2)\,dx$ because the $ikt$ term grows arbitrarily
rapidly in $t$ for $k$ arbitrarily large. The very simplest system
with this property occurs for $\lambda=0$, so that $\phi_{2,x}=0$ and
$\phi_{1,t}=\phi_{2,x}$.

Such systems are called ``weakly well-posed'', which means they are
not well-posed in $L^2$ but can be made well-posed if a higher
derivative norm is used for the initial data than for the solutions,
see \cite{KreissLorenz2004},~p.~39. In this example, we would include
$\phi_{2,x}^2$ in the norm at $t=0$ only, or we could introduce an
additional variable $\phi_3:=\phi_{2,x}$. They can become completely
ill-posed if lower-order terms are included, see
\cite{KreissLorenz2004},~p.~40 for an example.

We give this example to make two points: 1) when a plane-wave ansatz
does not give all the solutions of a linear PDE system with constant
coefficients, we should look for additional solutions that are
polynomial; 2) the problem with polynomial solutions is not that they
grow but that they can grow arbitrarily rapidly in time for initial
data that oscillate arbitrarily rapidly in space. (In our null toy
problem, read $x$ for ``time'' and $\bar y$ for ``space''.)


\section{Notes on \cite{Frittelli2005a}}
\label{appendix:Frittellicorrections}


Eq.~(\ref{Pabeqncorrected}) corrects
the right-hand side of (15a) of \cite{Frittelli2005a} by a factor of
$-1/2$. This error occurs already between the nonlinear field equation
(8) of \cite{Frittelli2005a} and its linearisation
(11a). 

Eq.~(\ref{Qaeqncorrected}) corrects the right-hand side of (15b) of
\cite{Frittelli2005a} by the addition of
$Q_a+2T_a$. Eq.~(\ref{Jaeqncorrected}) corrects the right-hand side of
(15e) of \cite{Frittelli2005a}, by the subtraction of
$Q_a+2T_a$. These errors occurs between (11c) and (15b) and (15e),
respectively. Note there is no error between (6b) and its
linearisation (11c).

The last two errors are related because the definitions (\ref{Pabdef})
and (\ref{Jadef}) give rise to the integrability condition
\begin{equation}
\nabla^bP_{ab}=\left(r\nabla^b\tilde
h_{ab}\right)_{,r}=\left(r(J_a-Q_a-2T_a)\right)_{,r},
\end{equation}
or, using (\ref{Taeqn}),
\begin{equation}
rQ_{a,r}+\nabla_bP_{ab}=rJ_{a,r}+J_a-Q_a-2T_a.
\end{equation}
Using the correct expression (\ref{Jaeqncorrected}) for $rJ_{a,r}$
then gives the correct equation (\ref{Qaeqncorrected}) for $Q_{a,r}$.

(\ref{Vtildeeqn}) is the covariant form of Eq.~(15c) of
\cite{Frittelli2005a}. A factor of 2 is missing from the last two terms
of Eq.~(12) of \cite{Frittelli2005a}, but is restored in Eq.~(15c).

A second set of minor corrections concerns the explicit matrices
$C^\mu$.  With $\phi^\dagger:=(P_{ab},Q_{a},...)$, Frittelli states
that $C^u=\text{diag}(1,1,0...,0)$ and
$C^r=\text{diag}(-{\cal A},1,1,...,1)$, but this cannot be true as stated
for any choice of normalisation of the equations. 

The matrices $C^a$ are not given explicitly, but become symmetric,
without terms $\sin^2\theta$ appearing in $C^u$ and $C^r$, only if one
introduces frame components on the 2-sphere as in (\ref{S2frame}). With the
variables in the order
\begin{equation}
\phi^\dagger=(P_{\theta\theta},P_{\theta{\hat\varphi}},Q_\theta,Q_{\hat\varphi},
{\cal P}_\theta,{\cal P}_{\hat\varphi},{\cal Q},\hat{\cal Q},...),
\end{equation}
where the dots stand for all other variables, in any order, then with
our normalisation of the equations the matrices are
\begin{eqnarray}
\label{Cudef}
C^u&=&\text{diag}(C^u_4,C^u_4,0,...,0), \\
\label{Crdef}
C^r&=&\text{diag}(C^r_4,C^r_4,1,...,1), \\
C^\theta&=&\text{diag}(C^\theta_4,C^\theta_4,0,...,0), \\
C^\varphi&=&{1\over\sin\theta}\text{diag}(C^{\hat\varphi}_4,C^{\hat\varphi}_4,0,...,0), 
\end{eqnarray}
where we have defined
\begin{eqnarray}
C^u_4&=&\text{diag}(2,2,0,0), \\
C^r_4&=&\text{diag}(-{\cal A},-{\cal A},1,1), \\
C^\theta_4&:=&{1\over r}\left(\begin{array}{cccc}
0 & 0 & 1 & 0 \\
0 & 0 & 0 & 1 \\
1 & 0 & 0 & 0 \\
0 & 1 & 0 & 0 \\
\end{array}\right) \\
C^{\hat\varphi}&:=&{1\over r}\left(\begin{array}{cccc}
0 & 0 & 0 & -1 \\
0 & 0 & 1 & 0 \\
0 & 1 & 0 & 0 \\
-1 & 0 & 0 & 0 \\
\end{array}\right).
\end{eqnarray}


\section{Symmetric hyperbolic form of the toy model}
\label{toymodelsymmetrichyperbolic}


To link the considerations of Sec.~\ref{section:hyperbolicity} and
Sec.~\ref{section:Frittelli}, in this Appendix we construct a
symmetric hyperbolic reduction of the toy model of
Sec.~\ref{section:hyperbolicity} in Bondi gauge, using the methods of
Sec.~\ref{section:Frittelli}.

We introduce the reduction variables 
\begin{eqnarray}
\label{Ptoydef}
P&:=& \overline{\delta f}_{,x}, \\
Q&:=& \overline{\delta b}_{,x}+\delta G_{,{\bar y}}, \\
\label{Jtoydef}
J&:=& \overline{\delta f}_{,{\bar y}}-\overline{\delta b}_{,x}, \\
\label{Ttoydef}
T&:=& \delta G_{,{\bar y}}, \\
\label{Ldef}
L&:=&J_{,y}-T_{,y}
\end{eqnarray}
Here $P$, $Q$, $J$ and $T$ are closely related to $P_{ab}$, $Q_a$,
$J_a$ and $T_a$ of Sec.~\ref{section:Frittelli}. They do not carry
indices because of the restriction to twist-free axisymmmetry. 

A symmetric hyperbolic first-order reduction of the system
(\ref{myGeqnB}-\ref{myfeqnB}) in terms of these variables is
\begin{eqnarray}
2P_{,u}-P_{,x}-Q_{,{\bar y}}&=&0, \label{Ptoyeqn} \\
Q_{,x}-P_{,{\bar y}}&=&0,\\
\overline{\delta b}_{,x}&=&Q-T, \\
J_{,x}&=& 0, \\
T_{,x}&= &0, \\
\overline{\delta f}_{,x}&= &P, \\
\delta G_{,x}&=& 0, \label{Gtoyeqn} \\
L_{,x}&=&0, \label{Ltoyeqn} \\
\overline{\delta H}_{,x}&=&L, \label{Htoyeqn}
\end{eqnarray}
in the variables
\begin{equation}
\phi^\dagger:=(P,Q,\overline{\delta
  b},J,T,\overline{\delta f},\delta G,L,\overline{\delta H}),
\end{equation}
The three matrices $C^\mu$ are symmetric and $C^t:=C^u+C^x$ is
positive definite. 

The bad solution $\overline{\delta b}=-{\delta G}_{0,{\bar y}}\,x$ of
the toy model becomes $\overline{\delta b}=-T_0x$ in the
symmetric-hyperbolic form, and the bad solution $\overline{\delta
  H}={\delta f}_{0,{\bar y}{\bar y}}\,x$ becomes $\overline{\delta
  H}=L_0x$. To see that the linear growth in $x$ is by itself in
conflict with our estimates, consider a toy model of the toy model,
namely
\begin{eqnarray}
T_{,x}&=&0, \\
b_{,x}&=&-T,
\end{eqnarray}
where for simplicity $T$ and $b$ depend only on $u$ and $x$. The
general solution is
\begin{eqnarray}
T&=&T_0(u), \\
b&=&b_0(u)-T_0(u)(x-x_0).
\end{eqnarray}
Comparing to our general framework for estimates, we see that $c=1$,
$d=1$, ${\bf Q}:=(T,b)$ and ${\bf P}$ is absent. Hence the estimate
(\ref{V1estimate}) reduces to 
\begin{equation}
\int_{t=t_1}{\bf Q}^\dagger{\bf Q}\le e^{t_1-t_0}\int_{x=x_0}{\bf Q}^\dagger{\bf Q}.
\end{equation}
We can use $u=t-x$, $u_1=t_1-x_0$, $u_0=t_0-x_0$ to write both
sides as integrals over $u$, namely 
\begin{eqnarray}
\int\displaylimits_{t=t_1}{\bf Q}^\dagger{\bf Q}&=&\int_{u_0}^{u_1}\left[T_0(u)^2+b_0(u)^2\right]\,du, \\
\int\displaylimits_{x=x_0}{\bf Q}^\dagger{\bf Q}&=&\int_{u_0}^{u_1}\Bigl[T_0(u)^2 \nonumber \\
&&+\left(b_0(u)-T_0(u)(u_1-u)\right)^2 \Bigr]\,du.
\end{eqnarray}
One can show that
\begin{eqnarray}
{\int\displaylimits_{t=t_1}{\bf Q}^\dagger{\bf Q} \over
\int\displaylimits_{x=x_0}{\bf Q}^\dagger{\bf Q}}&\le& 1+{\Delta^2\over 2}
\left(1+\sqrt{1+{\Delta^2\over 4}}\right), \\
&\le& e^{t_1-t_0} \text{ for } t_1\ge t_0,
\end{eqnarray}
where $\Delta:=u_1-u_0=t_1-t_0$, and the first inequality is sharp.
We should stress again that this is only a toy model, similar to the
one considered in Sec.~6.4 of \cite{Giannakopoulosthesis}.


\section{Attempts to extend the estimates to $r_0=0$}
\label{appendix:tricks}


We attempt to overcome the problem that on the Minkowski background
$d=\bar d/r$ with $\bar d>0$, which means that we cannot take $r_0\to
0$ in any of our estimates. In this Appendix we try three simple ways of
overcoming this, first by a change of integration measure $dV$, as in
Eq.~(\ref{omegaconservationlaw}).  The most general weight compatible
with the time translation and rotation symmetries of the Schwarzschild
or Minkowski background is
\begin{equation}
dV=du\,\omega(r)\,dr\,d\Omega,
\end{equation}
where $d\Omega=\sin\theta\,d\theta\,d\varphi$ is the integration
weight on $S^2$ induced by the round unit metric $q_{ab}$. The
choice $\omega(r)=r^2$ gives $dV$ induced by the background
Schwarzschild metric $g_{\mu\nu}$. 

From (\ref{omegaconservationlaw}) we see that both $d$ and $c$ acquire
a factor of $\omega$, so $\omega$ cancels out of the ratio $d/c$ that
appears in our estimates. This leaves us with the addition of $C^r$ to
play with. For simplicity, we make the ansatz $\omega(r)=r^p$, so that
(\ref{Somegadef}) becomes
\begin{equation}
S_\omega=S+{p\over r} j^r=\phi^\dagger\left({\cal D}+{p\over r}C^r\right)\phi.
\end{equation}
For the wave equation on Minkowski, we find that $\bar d=0$ for $p=1$
only, while it is positive for all other $p$. It is also positive on
Schwarzschild for all $r<\infty$, for all $p$. So for the wave
equation on Minkowski only, the unique choice $\omega=r$ gives us
$d=0$. 

For the linearised Einstein equations on the Minkowski background, we
find that the largest eigenvalue $\bar d$ of $r{\cal D}+p C^r$ has a
minimum of $\bar d \simeq 3.4$ at $p\simeq -2.2$, that is, we cannot
make $\bar d$ non-positive with any choice of $p$. Hence we cannot
make $\bar d$ non-positive in this way.

For a second attempt, we observe that, with $\tilde\beta_{,r}=0$ and
$\tilde\beta=0$ at regular centre, $\beta$ and its angular derivatives $T_a$
and ${\cal T}$ vanish identically, as they must vanish at the
centre. If we experimentally take them out of the system, then $\bar
d\simeq 3.1$. If we additionally bring $\omega(r)=r^p$ into play, $\bar d$
has a minimum of $\bar d \simeq 2.6$ at $p\simeq -1.2$. Again, this is
not enough to make $\bar d$ non-positive.

For a third attempt, we attempt to change the ratio $d/c$ by a linear
recombination of variables as in Eq.~(\ref{phitildedef}). We have already seen
that the eigenvalues of ${\cal D}$ and $C^t$ can change if $A$ is not
orthogonal. For the full system we cannot try out all possible
matrices $A$ by brute force, but we can for the scalar wave equation
on Minkowski, as we shall see now. 

From (\ref{waveeqnPthetahatvarphi}-\ref{waveeqnpsicopy}) we read off
that on Minkowski $C^t:=C^u+C^r$ is the unit matrix, while
\begin{equation}
{\cal D}=D+D^\dagger={1\over r}\left(\begin{array}{cccc}
0 & 0 & 0 & 1 \\
0 & -2 & 0 & 0 \\
0 & 0 & -2 & 0 \\
1 & 0 & 0 & -2 \\
\end{array}\right).
\end{equation}
We have $\bar d=\sqrt{2}-1\simeq 0.4$. We have also just seen that
with $p=1$ we can make this $\bar d=0$. However, we now focus on
linear recombinations of the variables other than by an overall
$r$-dependent factor.

Geometrically, it makes no sense to mix the components of the vector
$Q_a$ either with each other or with the scalars $P$ and $\psi$,
and we can fix an overall factor in $A$ by leaving them completely
unchanged, so $A$ acts non-trivially only on the pair
$(P,\psi)$. Hence we can assume
\begin{equation}
A=\left(\begin{array}{cccc}
\alpha & 0 & 0 & \beta \\
0 & 1 & 0 & 0 \\
0 & 0 & 1 & 0 \\
\gamma & 0 & 0 & \delta \\
\end{array}\right).
\end{equation}
Two of the eigenvalues $\tilde {\cal D}:=A^\dagger{\cal D}A$ are then
always $-2/r$ for any $A$. As they are negative, this is not a
problem. Furthermore, if we write $A=\bar A R$ where $R$ is
orthogonal, $R$ does not change the eigenvalues of $\tilde C^t$ and
$\tilde{\cal D}$. Hence we can choose the rotation $R$ between $P$ and
$\psi$ such that
$\gamma=0$, and without loss of generality we can then rescale $\psi$ to
set $\delta=1$.  The non-trivial eigenvalues of $\tilde {\cal D}$ are now
$(\beta-1)\pm\sqrt{(\beta-1)^2+\alpha^2}$. We have already used up the rotation
between $P$ and $\psi$ that would allows us to set $\alpha=0$, and hence
one of the eigenvalues of $\tilde{\cal D}$ is always positive.

In summary, none of three relatively trivial ways of changing $d/c$ in
the estimate manage allows us to make it non-positive, and as $d/c$ is always
proportional to $1/r$, we cannot let $r_0\to 0$ in estimates based on
our symmetric hyperbolic reduction.


\section{Numerical experiments}
\label{appendix:numericalexperiments}


We have not been able to obtain an estimate (for the linearised
Einstein equations) for the PDE problems we have been solving
numerically in \cite{axinull_formulation,axinull_critscalar}, where
our time slices are outgoing null cones {\em emanating from a regular
  centre}.  The reason for this was that the factor $\exp
d(r_0)\,(t_1-t_0)$ in our estimates diverges as $r_0\to 0$. However,
we did not actually notice an instability in our code. In this
appendix, we test the code a little harder by running it with
small amplitude but random initial data, and see if there is a
discrete norm that remains bounded in by its initial value. The
control volume is as in Fig.~\ref{fig:V3}, but with $r_0=0$,
understood as a regular centre.

We now construct a hypothesis to test by modifying the estimate
(\ref{V3estimate}) as follows. $r=0$ is not actually a boundary, but
an interior world line, and no free data can be imposed there, so the
term at $r=r_0$ disappears. In the setup of our code we are not
interested in controlling the energy leaving the control volume
through $v=v_1$, so we drop that term from the estimate
(\ref{V3estimate}). The hypothesis we want to test is that the function
norms on $u=0$ and $u=u_1$ are the ones that we derived but that an
unknown function $K(u_1-u_0)$ takes the place of $\exp
d(r_0)\,(u_1-u_0)$. Hence our hypothesis is
\begin{equation}
\int\displaylimits_{u=u_1}{\bf P}^\dagger{\bf P}
\le K(u_1-u_0) \int\displaylimits_{u=u_0}{\bf P}^\dagger{\bf P},
\label{V4estimate}
\end{equation}
where ${\bf P}^\dagger{\bf P}$ was defined above in
(\ref{PdaggerPdef}).  

Restricting to vacuum in twist-free axisymmetry,
and translated into the notation of our code, the two terms in
${\bf P}^\dagger{\bf P}$ are
\begin{eqnarray}
{1\over 2}P^{ab}P_{ab}&:=&{1\over 2}q^{ac}q^{bd}
(r\tilde h_{ab})_{,r}(r\tilde h_{cd})_{,r} \\
&=&(r\tilde h_{\theta\theta})_{,r}^2 \\
&=&4S^2(rf)_{,r}^2 \label{integrandterm1}
\end{eqnarray}
and
\begin{eqnarray}
P^aP_a&:=&q^{ab}(r\nabla^c\tilde h_{ac})_{,r}
(r\nabla^d\tilde h_{bd})_{,r} \\
&=&[r(\tilde h_{\theta\theta,\theta}+2\cot\theta \,\tilde h_{\theta\theta})]_{,r}^2\\
&=&4S[S(rf)_{,ry}-8y(rf)_{,r}]^2.
\end{eqnarray}
(Recall $y:=-\cos\theta$ and $S:=1-y^2$.) Note, however, that $\delta H$ forms
part of the variables ${\bf Q}$, that these are not coming in through
$r=0$, and we do not control their leaving through $v=v_1$. Hence
instead of our full estimate we can use Frittelli's truncated
estimate, with ony the term (\ref{integrandterm1}) in the integrand.

In the linearised equations, the scalar matter field decouples from
the metric perturbations, and we conjecture that it obeys a similar
estimate with integrand
\begin{equation}
P^2:=(r\psi)_{,r}^2.
\end{equation}

Our integration measure in axisymmetry is 
\begin{equation}
\int dS=\int d\Omega\,dR=2\pi \int_{-1}^1dy\,\int_0^{x_{\max}}R_{,x}\,dx,
\end{equation}
and we now drop the factor $2\pi$. Our code requires the outer
boundary to be null or future spacelike, so we cannot literally run in
Bondi gauge $R=x$. However, runnning in lsB gauge and setting
$x_0=x_\text{max}$ will turn the outer boundary $x=x_\text{max}$ into
an ingoing null In weak gravity, this will result in $R=R(u,x)\simeq
c(u)x$, with $c(u)$ determined so that $x=x_0$ is ingoing null. We can
then identify $x=x_0$ with $v=v_1$. See \cite{axinull_formulation} for
details.

To test our conjecture, we therefore evaluate the norms (at time $u$)
defined by
\begin{eqnarray}
 ||(r\psi)_{,r}||^2&:=&\int_{-1}^1dy \int_0^{x_{\max}}
  {[(R\psi)_{,x}]^2\over R_{,x}}\,dx, \\
||(rf)_{,r}||_1^2&:=&\int_{-1}^1 (1-y^2)\,dy \int...f...
\end{eqnarray}
and their discretizations
\begin{eqnarray}
\label{drpsidrnorm}
||(r\psi)_{,r}||_1^2&\simeq& 2 \sum_{j=1}^{N_y}A^{(0)}_{l=0,j}
\sum_{i=1}^{N_x}
{\left(R_{i}\psi_{i,j}-R_{i-1}\psi_{i-1,j}\right)^2\over
  R_{i}-R_{i-1}},  \nonumber \\ \\
\label{drfdrnorm}
||(rf)_{,r}||_1^2&\simeq& 2\sum A^{(0)}_{l=0,j}(1-y_j^2)\sum...f..., 
\end{eqnarray}
where $R_i$ has only one index as $R=R(u,x)$ in lsB gauge as $R$ does
not depend on $y$. We have also used that $1/2\int \psi dy$ is the
$l=0$ component of $\psi$, and have used the analysis matrix $A^{(0)}$
of our pseudospectral framework to determine it. Our radial grid
starts at $i=1$, but in the formulas we use $R=0$ at $i=0$. See again
\cite{axinull_formulation} for details.

We have evolved with noise of amplitude $10^{-10}$ in $f$ and $\psi$,
at $N_x=64...1024$ and $N_y=17...65$. We set $x_\text{max}=1.1$ and
$x_0=1$, and evolve to $u=0.95$, when the range of $R$ has gone down
from its initial value of $0.5$ by a factor of $1-0.95=0.05$. We do
not filter out high frequencies, other than the components of $f$ with
$l=N_x+1$ and $N_x$ (see \cite{axinull_formulation} for why we do
this.) At each $x_i$, we also set all spherical harmonic components
with $l>2i$ to zero (see again \cite{axinull_formulation} for why we do
this.)

The discrete $L^2$ norms of $\psi$ and $f$ themselves do not decrease
with $u$, but the discretised $L^2$ norms $||(r\psi)_{,r}||$ and
$||(rf)_{,r}||_1$ given in (\ref{drpsidrnorm},\ref{drfdrnorm})
do. From about $u=0.01$, both decrease
approximately as $u^{-0.35}$. Near the end, $||(r\psi)_{,r}||$
decreases much more rapidly, while $||(rf)_{,r}||$ increases a bit
in an apparently random manor, then decreases again. In summary, our
numerical experiments are easily compatible with the estimate
(\ref{V4estimate}) for $K(u_1-u_0)=1$ and including only $P^{ab}P_{ab}$,
that is
\begin{equation}
\int\displaylimits_{u=u_1}(rf)_{,r}^2\,dr\,dy 
\le \int\displaylimits_{u=u_0}(rf)_{,r}^2\,dr\,dy,
\label{V4estimatebis}
\end{equation}
and a similar estimate with $\psi$ instead of $f$. 

Looking at individual spherical harmonic components $f_l(u,x)$ and
$\psi_l(u,x)$ of $f(u,x,y)$ and $\psi(u,x,y)$, we see that the random
initial dominated by the grid frequency data quickly smooth out into
well-resolved data on frequencies lower than the grid frequency. Hence
our method seems to be quite dissipative. In a second phase we then
see that these appear to be ``stretched'' in $x$, as the grid in fact
zooms in on them, with $R_\text{max}$ shrinking linearly in $u$.

For definiteness, we have implemented a specific continuum norm and
its discretisation, but we should stress that this was just an
informed guess. In particular, we have arbitrarily chosen
$dV=d\Omega\,dR$, rather than, for example, $dV=d\Omega\,R^p\,dR$. The
``correct'' choice would depend on what (if any) estimate can be
proved. We have also not gone to particularly large values of $N_x$
and $N_y$, but only ones that we have also used in physics
simulations. (A soft upper limit on $N_x$ is set by computing time,
while a hard limit $N_y\le 128$ is set by the accuracy of our spectral
method in $y$.) Hence our results should only be considered as a
slightly more challenging stability test of our code motivated by the
results of this paper, not a numerical test of well-posedness.



\end{document}